\documentclass{pas}
\usepackage{multirow}
\usepackage{hyperref}
\usepackage{amsmath}
\usepackage{makecell}
\usepackage{orcidlink}
\usepackage[pagewise]{lineno}

\begin{document}

\lefttitle{Publications of the Astronomical Society of Australia}
\righttitle{S. Panjkov \textit{et al.}}

\jnlPage{1}{17}
\jnlDoiYr{2026}
\doival{10.1017/pasa.2026.10228}

\articletitt{Research Article}

\title{Estimating the Local Star Formation Rate Density from ASKAP RACS}

\author{Sonja Panjkov$^{1}$\orcidlink{0000-0002-1719-2024}, O. Ivy Wong$^{2,3}$\orcidlink{0000-0003-4264-3509} and Rachel L. Webster$^{1}$\orcidlink{0000-0002-5325-2709}}

\affil{$^1$School of Physics, The University of Melbourne, Parkville, VIC 3010, Australia, $^2$Space $\&$ Astronomy, Commonwealth Scientific and Industrial Research Organisation (CSIRO), Bentley, WA, Australia, $^3$International Centre for Radio Astronomy Research, The University of Western Australia, Crawley, WA, Australia}

\corresp{S. Panjkov, Email: srpanjkov@gmail.com}

\citeauth{Panjkov S,  Wong OI and Webster RL, Estimating the Local Star Formation Rate Density from ASKAP RACS. {\it Publications of the Astronomical Society of Australia} {\bf 43}, e094, 1-17. https://doi.org/10.1017/pasa.2026.10228}

\history{(Received 07 December 2025; revised 14 May 2026; accepted 18 June 2026)}

\begin{abstract}

Understanding the evolution of the cosmic star formation rate density (SFRD) is key to uncovering how the Universe arrived at its present state. This paper presents a novel and efficient method to estimate the local SFRD, which uses supervised machine learning to first identify a population of star-forming galaxies (SFGs). Next, star-formation rates (SFRs) are determined using the 1.4-GHz radio-continuum emission detected by the Australian Square Kilometre Array Pathfinder (ASKAP). Specifically, a gradient-boosted decision tree model was implemented to classify extragalactic sources from the \citet{Beck2022} catalogue as either galaxies or quasars using RACS-mid and WISE photometry. The full sample, consisting of 389,392 sources, was partitioned into a 70\%-15\%-15\% split for training, validating, and testing. The optimised model achieved a weighted F1 score of 0.93 and an accuracy of 0.94 on the test dataset, ultimately classifying 336,674 sources as galaxies and 52,718 sources as quasars. Using the resulting $z<0.1$ depth-matched galaxy sample and the photometric redshift predictions from \citet{Beck2022}, a modified 1.4-GHz SFR calibration was determined, yielding a local, completeness-corrected, $z<0.1$ SFRD of $(1.4 \pm 0.5) \times 10^{-2} \; \rm M_{\odot}\, yr^{-1}\,Mpc^{-3}$ using 11,293 sources. This value is consistent with previous results. Thus, this study demonstrates the feasibility of using supervised learning to identify large populations of SFGs in order to investigate the SFRD evolution. This presents an exciting prospect for future, deeper surveys such as EMU, which will enable the cosmic SFRD to be probed out to higher redshifts.

\end{abstract}


\maketitle

\section{Introduction}

In recent years, machine learning has proven successful in the classification of both galactic and extragalactic sources \citep[e.g.][]{Bates2012, Lyon2016, Tan2018, Liu2019, Karsten2023, Riggi2024}. For example, \citet{Beck2021} and \citet{Beck2022} implemented deep neural networks to separate populations of galaxies, quasars, and stars using Pan-STARRS1 optical and WISE IR photometry, achieving galaxy completeness and purity scores of 99.25\% and 99.15\%. Importantly, many of these machine-learning implementations were not trained on spectroscopic data, but instead relied on published photometric catalogues that often contain $\gtrsim 10^6 - 10^9$ sources, providing a quick and cheap alternative to classify large, astrophysical populations.

In this paper, we propose and test the viability of a new method to constrain the SFRD which uses supervised learning to classify populations of galaxies and quasars using readily available radio continuum and IR catalogues. Since 1.4 GHz radio emission is a reliable tracer of star-formation activity for galaxies whose radio emission isn't dominated by AGN activity, galaxies selected in this manner can be used to investigate the physics of star formation. This allows us to estimate the local ($z<0.1$) SFRD using a sample of 11,293 machine-learning selected galaxies and a modified 1.4-GHz SFR calibration that was obtained following the methodology of \citet{Molnar2021}. 

This paper is structured as follows. In Section \ref{sec:data}, we describe the data used as inputs and labels for the model. In Section \ref{sec:methods}, we outline the model architecture and training approach, and in Section \ref{sec:results}, we assess the performance of the model. In Section \ref{sec:sfrs}, we estimate the local SFRD using the 1.4 GHz emission of the $z<0.1$ depth-matched galaxy sample, and correct this result for completeness in Section \ref{sec:completeness_corr}. We discuss the efficacy and the limitations of the method in Section \ref{sec:sfrd} and conclude in Section \ref{sec:conc}. Throughout the paper, we adopt a $\Lambda$ cold dark matter ($\Lambda$CDM) cosmology with $\Omega_M=0.27$, $\Omega_{\Lambda}=0.73$, $\Omega_K=0$, and $H_0=73 \rm \;km\,s^{-1} \,Mpc^{-1}$.

\section{Datasets}\label{sec:data}

\subsection{RACS-mid}

The Australian Square Kilometre Array Pathfinder (ASKAP, \citealt{Hotan2021}) has surveyed $\sim90\%$ of the sky as part of the Rapid ASKAP Continuum Survey \citep[RACS,][]{McConnell2020}, with the aim of providing a sky model for the calibration and validation of future ASKAP surveys. RACS commenced in 2019 and observed each area of the sky with a $\sim$15-minute integration time, achieving a root-mean-square noise of $\sim 200 - 260 \, \rm \mu Jy \, PSF^{-1}$ and a resolution of $\sim8 - 15$ arcseconds. RACS consisted of three major observing epochs, RACS-low \citep{McConnell2020, Hale2021}, RACS-mid \citep{Duchesne2024}, and RACS-high \citep{Duchesne2025}, centered on effective frequencies of 887.5, 1367.5 MHz, and 1655.5 MHz, respectively.

Since radio continuum emission has the potential to trace SFRs without significant dust extinction \citep{Gordon2004, Osterbrock1989, Bowler2015}, tabular data from the RACS-mid catalogue are used as features for the binary-classification model implemented in this paper. The second data release of RACS-mid \citep{Duchesne2024} contains 1493 sky images at 1367.5 MHz with a bandwidth of 144 MHz. These images were convolved to a common resolution and mosaicked, yielding a series of near-constant sensitivity images that cover the sky up to a declination of $+49^{\circ}$. Accompanying these images are source and component catalogues, consisting of 3,105,668 radio sources with an estimated completeness of 95\% above 2mJy \citep{Duchesne2024}. 

The source catalogue was accessed through the Commonwealth Scientific and Industrial Research Organisation (CSIRO) ASKAP Science Data Archive\footnote{The CSIRO ASKAP Data Science Archive can be accessed at https://research.csiro.au/casda/.}. Source features in the catalogue include the peak and total flux densities; the position angle and major and minor axes of the source, the deconvolved source, and the full width at half maximum of the PSF of the field; the number of Gaussian components of the source and the source structure code (single or multiple Gaussians); and the source type flag (resolved, unresolved, or spurious).

\subsection{WISE All-Sky}

NASA's Wide-Field Infrared Survey Explorer (WISE, \citealt{Wright2010}) imaged the full infrared sky across four pass bands centred at 3.4, 4.6, 12, and 22 $\mu$m (W1, W2, W3, W4), achieving 5$\sigma$ point source sensitivities greater than 0.08, 0.11, 1, and 6 mJy respectively in uncrowded regions. Following the depletion of its coolant, WISE continued mapping the sky in the W1 and W2 passbands with the aim of detecting near-Earth moving objects including asteroids and comets, as the Near-Earth Object Wide-field Infrared Survey Explorer (NEOWISE, \citealt{Mainzer2011, Mainzer2014}) until the mission ended in 2024. 

The WISE All-Sky Catalogue \citep{Cutri2012} contains IR photometry and astrometry for over 300 million sources, and was combined with data collected during NEOWISE to produce the AllWISE catalogue \citep{cutri2021}, which includes $\sim750$ million sources and improved sensitivity in the W1 and W2 bands. unWISE is the most recent and complete release of WISE and NEOWISE data, with  $\sim2$ billion sources thanks to deeper imaging and improved modelling of crowded regions \citep{Schlafly2019}. However, in line with the model and data used by \citet{Beck2022}, data fields from the WISE All-Sky Catalogue are used as inputs for the binary-classifier model. The source catalogue was accessed via the NASA/IPAC Infrared Science Archive (IRSA)\footnote{https://irsa.ipac.caltech.edu/}.

\subsection{WISE-PS1-STRM}

The WISE $\times$ PanSTARRS1 Source Types and Redshifts with Machine learning (WISE-PS1-STRM, \citealt{Beck2022}) catalogue contains source classifications and photometric redshifts for 354,590,570 objects. Building on an earlier catalogue that classified PanSTARRS1 3$\pi$ Data-Release 1 (PS1 3$\pi$ DR1, \citealt{Chambers2016, Flewelling2020}) sources using supervised learning (PS1-STRM, \citealt{Beck2021}), the WISE-PS1-STRM catalogue was constructed by cross-matching WISE All-Sky and PS1 $3\pi$ DR2 sources. The cross-matched objects were then classified into galaxies, quasars, and stars using a deep neural network, with the inclusion of WISE IR colours achieving greater star sample purity and quasar sample completeness compared to the earlier model which relied on optical PS1 observations alone. 

The training set for the \citet{Beck2022} machine learning algorithm consisted of matched spectroscopic observations, mainly from the 14th data release of the Sloan Digital Sky Survey \citep[SDSS DR14][]{Abolfathi2018}, but supplemented by spectra from the fourth data release of the DEEP2 Redshift Survey \citep{Newman2013}, the VIMOS Public Extragalactic Redshift Survey (VIPERS) public data release 2 (VIPERS PDR-2, \citealt{Scodeggio2018}), the WiggleZ Dark Energy Survey \citep{Drinkwater2018}, the 3rd data release of zCOSMOS \citep{Lilly2009}, and the VIMOS VLT Deep Survey (VVDS, \citealt{LeFevre2013}). A separate deep neural net was then used to estimate photometric redshifts for those objects they classified as galaxies. 

For the supervised-learning model implemented in this paper, the source classifications from WISE-PS1-STRM are used to label the training data set of cross-matched RACS-mid and WISE All-Sky sources. WISE-PS1-STRM data was accessed via Mikulski Archive for Space Telescopes (MAST) CasJobs\footnote{https://mastweb.stsci.edu/mcasjobs/}.

\section{Supervised-Learning Algorithm Implementation}\label{sec:methods}

This section describes the datasets and algorithms that were used to train a simple binary classification model to predict the natures of the RACS-mid sources using the galaxy and quasar labels from \citet{Beck2022}. These model predictions are then used to select an uncontaminated sample of SFGs that enable the local SFRD to be estimated in Section \ref{sec:sfrd}.

\subsection{Our Dataset}

To obtain the data and labels for the binary-classifier model, the RACS-mid catalogue from \citet{Duchesne2024} was cross-matched with the WISE-PS1-STRM catalogue from \citet{Beck2022} using TOPCAT version 4.10-2\footnote{http://www.starlink.ac.uk/topcat/} \citep{Taylor2005}. The cross-match was performed using the sky algorithm with a maximum error of 1.5 arcseconds, ensuring that no row from either table appeared more than once in the final table, yielding an initial cross-match sample consisting of 426,815 sources. Any sources that were classified as stars or were listed as unsure in the \citet{Beck2022} catalogue were then dropped. The final cross-match sample consisted of 389,392 sources, of which 14.2\% were quasars and 85.8\% were galaxies. 

\subsection{Our Model}

To classify the final cross-matched sample into galaxies and quasars, a gradient-boosted decision tree model from the \texttt{XGBoost} package \citep{Chen2016} was implemented. Here, we describe our approach for this model.

\subsubsection{Input Features}

The same WISE features used by \citet{Beck2022} were selected as model inputs. These included the profile-fit photometry and uncertainties in the four WISE passbands for different aperture sizes, as well as features relating to the observation quality. In total, this amounted to 68 WISE features.

For the RACS-mid data, all features that related to the source properties and the associated uncertainties were selected as model inputs, amounting to 28 features in total. These included the peak and total fluxes, extended sources flags, and features relating to the source size and extent. Details on the RACS and WISE features used as model inputs can be found in Table \ref{tab:feature list} in Appendix \ref{sec:features}.

Any qualitative features were assigned numerical values, and features with information relating to multiple bands were split into separate inputs, yielding 105 features in total. Details on this process can be found in Appendix \ref{sec:feature_engineering}.

\subsubsection{Training Approach}

To optimise and assess the performance of the binary-classifier model, the cross-matched sample was split into three sets for training, validation, and testing, each making up 70\%, 15\%, and 15\% of the total sample, respectively. These splits were performed using \texttt{StratifiedShuffleSplit} from the \texttt{Scikit-learn} package \citep{Pedregosa2011} to maintain the class representation of galaxies and quasars across each of the three groups. 

\begin{table}[]
\centering
\caption{The class distribution for the final training, validation and test datasets, including the contribution from real and synthetic sources after random oversampling of the quasar class.}
\begin{tabular*}{0.48\textwidth}{@{\extracolsep{\fill}} l l c c }
\toprule \toprule
Sample & Training & Validation & Testing \\
\toprule
Real Galaxies & 233743 & 50088 & 50088 \\
Real Quasars & 38831 & 8321 & 8321 \\
Synthetic Galaxies & 0 & 0 & 0 \\
Synthetic Quasars & 194912 & 0 & 0\\ 
Fraction Galaxies & 0.50 & 0.86 & 0.86\\
Fraction Quasars & 0.50 & 0.14 & 0.14\\
\toprule
\end{tabular*}
\label{tab:class_dist}
\end{table}

Given the significant class imbalance of the cross-matched sample ($\sim$85\% galaxies and $\sim$ 15\% quasars), the effects of various under- and over-sampling techniques were trialled using the \texttt{imblearn} package \citep{Lemaitre2017}. Resampling techniques that were tested included random oversampling (\texttt{RandomOverSampler}), random undersampling (\texttt{RandomUnderSampler}), oversampling using the Synthetic Minority Over-sampling Technique (\texttt{SMOTE}, \citealt{Chawla2002}), cleaning using Tomek links (\texttt{Tomek}, \citealt{Tomek1976}), and a combination of both SMOTE and Tomek links (\texttt{SMOTETomek}, \citealt{Batista2003}). Each of these resampling methods resulted in an even class distribution between galaxies and quasars, allowing the model to more effectively learn the patterns in the data. 

To save on computation time, a smaller training and validation set were used to assess the performance of the different resampling techniques prior to extending the pipeline to the full sample. This smaller data set consisted of cross-matched sources with $25^\circ < \delta < 50^\circ$, amounting to 58,389 sources, of which 49,452 were galaxies and 8,937 were quasars. 70\% of these sources were allocated for model training, while the remaining 30\% formed the test set. Note that these resampling methods were applied to the training data set only to avoid data leakage.  

For each resampling method, the mean F1 score across five cross-validation folds was calculated. Based on this, random oversampling of the minority class was found to achieve the best performance on the trial dataset. The same pipeline was therefore extended to the full, final cross-matched sample, consisting of 389,392 objects, of which $\sim86\%$ were galaxies and $\sim14\%$ were quasars. After oversampling, an even split was achieved between the two classes in the training set. Details of the final sample distribution between galaxies and quasars can be found in Table \ref{tab:class_dist}. The performance of this final model is discussed in Section \ref{sec:results}.

Hyperparameters were optimised using \texttt{RandomizedSearchCV} and \texttt{Hyperopt} following the process described in Appendix \ref{ref:hyperparams}, and \texttt{early\_stopping\_rounds} was set to 100 to avoid overfitting. The optimised model hyperparameters are listed in Table \ref{tab:optimised_params}. 

\section{Classification Results} \label{sec:results}

The final model performance was evaluated using the test data set and a series of metrics including accuracy, precision, recall, the F1 score, the cross-entropy, the area under the receiver operating characteristic curve (ROC AUC), and the area under the precision-recall curve (AUPRC). The metric values for the optimised model are provided in Table \ref{tab:optimised_metrics} and the confusion matrix for the model predictions on the full dataset is shown in Figure \ref{fig:confusion_matrix}. 

Model predictions were generated for the entire final cross-matched sample. Of the 389,392 sources, the model predicted that 336,674 are galaxies and 52,718 are quasars, making up $\sim86\%$ and $14\%$ of the sample, respectively. 

To provide an independent constraint on the AGN contamination fraction of the model-selected galaxy sample, we cross-matched those sources identified by the model as galaxies with the Milliquas catalogue \citep{Flesch2023}. This process identified 7712 AGN in our galaxy sample of the 152,060 galaxies in the overlapping region between the two catalogues. This corresponds to an AGN contamination fraction of 0.05 across all redshifts, or 0.03 for the subset of sources with $z<0.1$.


\section{Estimating the SFRs of the Model-Selected Galaxy Sample}\label{sec:sfrs}

\subsection{Radio and Total Infrared Luminosities of the Galaxy Sample} \label{sec:lums}

Having identified a population of galaxies using the optimised model, it is possible to estimate the SFRs using their 1.4 GHz radio-continuum emission, since the emission in this band is a reliable tracer of star formation that does not suffer significantly due to dust extinction \citep{Osterbrock1989, Gordon2004, Bowler2015, Molnar2021}. This and Section \ref{sec:depth_matching} below follow the methodology presented in \citet{Molnar2021} to determine a modified 1.4-GHz SFR  prescription for a sample of galaxies in a given redshift range. Note that since we use the predicted galaxy redshifts from \citet{Beck2021}, our luminosity and SFRD calculation below is limited to their galaxy sample. However, we do not include those galaxies that were missed by our model and classified as quasars.

The luminosity at frequency $\nu$ is given by:

\begin{multline}\label{eq:lum}
    \rm \bigg( \frac{L_{\nu}}{W \, Hz^{-1}} \bigg) = (9.52 \times 10^{15}) \times 4\pi \times K(z) \times \\ \rm \bigg(\frac{D_L}{Mpc} \bigg)^2 \bigg( \frac{S_{int}^{\nu}}{mJy}\bigg),
\end{multline}

\begin{table}[]
\centering
\caption{Performance metrics for the optimised \texttt{XGBClassifier} model with random oversampling of the quasar class.}
\begin{tabular*}{0.35\textwidth}{@{\extracolsep{\fill}} l c }
\toprule \toprule
Metric & Value  \\
\toprule
accuracy & 0.94 \\
weighted F1 score & 0.93 \\ 
quasar precision & 0.80 \\
quasar recall & 0.73 \\
quasar F1 score & 0.76 \\
galaxy precision & 0.96 \\
galaxy recall &  0.97 \\
galaxy F1 score & 0.96 \\
cross-entropy & 0.16 \\
ROC AUC & 0.97 \\
AUPRC & 0.99\\
\toprule
\end{tabular*}
\label{tab:optimised_metrics}
\end{table}

\noindent{where $9.52 \times 10^{15}$ is the conversion factor from $\rm Mpc^2 \, mJy$ to $\rm W\,Hz^{-1}$, $D_L$ is the luminosity distance to the source, and $S_{int}^{\nu}$ is the measured flux density at frequency $\nu$. The k-correction factor $\rm K(z) = (1 + z)^{-(1 + \alpha)}$ accounts for redshift-related effects and depends on the spectral index, $\alpha$, defined as $S_{int} \propto \nu^{\alpha}$. To calculate the 1.4-GHz radio luminosity, a radio spectral index of $-0.7$ is assumed, as is standard for SFGs at 1.4 GHz \citep{Kimball2008, Molnar2021}. Note that $S_{int}$ is measured at 1367.5 MHz for RACS-mid, however it was treated as equivalent to the 1.4 GHz integrated flux density for the purposes of this study.

\begin{figure}
    \centering
    \includegraphics[width=0.99\linewidth]{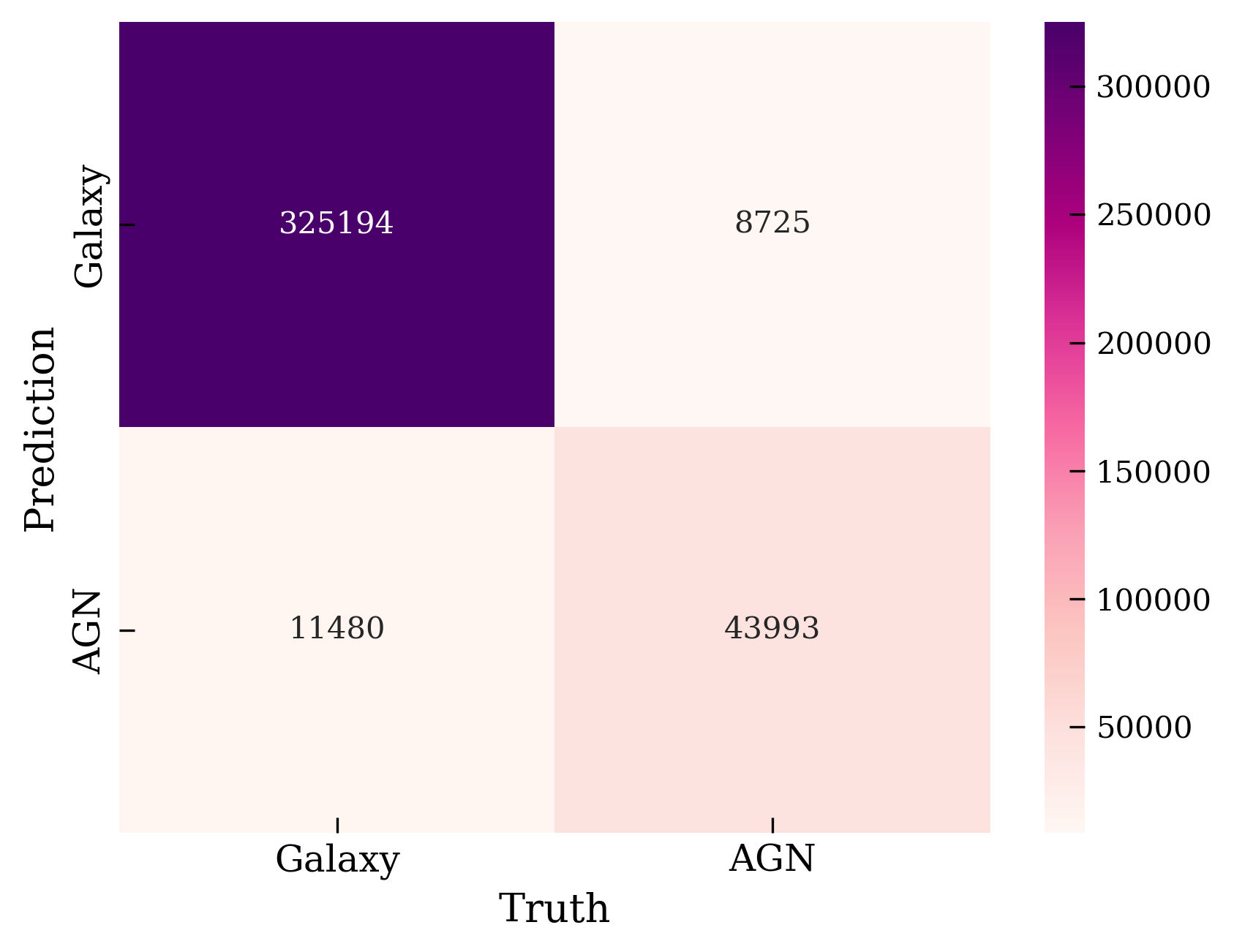}
    \caption{Confusion matrix for the model predictions for all 389,392 sources in the cross-matched dataset.}
    \label{fig:confusion_matrix}
\end{figure}

To calculate the total IR (TIR) luminosity, the observed WISE Vega magnitudes, $\rm m_{vega, x}$, are first converted to integrated flux densities, $\rm S_{int}^{Wx}$, via:

\begin{equation} 
    \rm S_{int}^{Wx} = S_{0, x} \times 10^{\frac{-m_{vega, x}}{2.5}},
\end{equation}

\noindent{where $x$ represents the WISE band number, and $\rm S_{0,x }$ is the zero magnitude flux density in the given band, with values of 390.54, 171.787, 31.674, and 8.363 for bands W1, W2, W3, and W4, respectively \citep{Wright2010, Jarrett2011}.}

Next, the methodology presented in \citet{Grundy2023} that is based on \citet{Cluver2017} is followed to determine the TIR luminosity. For SFGs, the WISE W3 band contains a significant contribution from the ISM, in addition to emission from evolved stellar populations. The ISM contribution arises from a variety of sources, but notably from polycyclic aromatic hydrocarbon (PAHs), which have been shown to trace regions of active star formation without suffering from dust-extinction biases \citep{Houck2007, Alonso-Herrero2014, Shipley2016, Kim2024b, Ronayne2024}. To remove the stellar contribution from the W3 band integrated flux, it is assumed as in \citet{Helou2004}, \citet{Cluver2017}, and \citet{Grundy2023}, that the near-infrared W1 band is a proxy for the stellar emission, and that 15.8\% of the W1 band stellar emission is also detected in the W3 band. Therefore, the W3PAH integrated flux component is obtained by subtracting 15.8\% of the W1 band flux from the W3 band flux. The W3PAH luminosity, $L_{W3PAH}$, is then calculated using Equation \ref{eq:lum} above, assuming a flat ($\alpha = 0$) IR spectrum for the k-correction factor \citep{Wright2010, Jarrett2011, Izotova2019}.

The TIR luminosity is finally obtained according to the relationship presented in \citet{Cluver2017}:

\begin{multline} \label{eq:lum_w3pah}
   \rm log\bigg( \frac{L_{TIR}}{L_{\odot}} \bigg) = (0.889 \pm 0.018) \, log \bigg( \frac{\nu L_{W3PAH}}{L_{\odot}} \bigg) \\ + (2.21 \pm 0.15),
\end{multline}

\noindent{where $\nu$ is the W3 band central frequency ($2.498 \times 10^{13}$ Hz, \citealt{Wright2010}).}

Figure \ref{fig:lradio_ltir_subplots} shows the Infrared-Radio Correlation (IRRC, 1.4 GHz vs. TIR luminosities) for the model-selected galaxy sample in different redshift bins. On a log-log scale, the slope of the best-fit linear model of the form $\rm log(L_{1.4 \,GHz}) = m\,log(L_{TIR}) +c$ indicates whether the IRRC is linear ($m=1$) or non-linear ($m\ne1$) across a given redshift range. In each panel of Figure \ref{fig:lradio_ltir_subplots}, the colour bar indicates the logarithm of the number of sources in each pixel, and the dashed blue line is the best-fit model ($m=1.114 \pm 0.009$, $c=10.2 \pm0.1$) for the depth-matched, $z<0.2$ galaxy sample from \citet{Molnar2021}. The black line shows the best-fit linear model to each data set, which was obtained using the \texttt{curve\_fit} command from the \texttt{scipy.optimize} package \citep{Virtanen2020}. The $1\sigma$ uncertainties in the linear-fit parameters were calculated using a bootstrapping technique, whereby the data points were resampled with replacement and a new linear fit was calculated for each iteration. The reported uncertainties are the standard deviations of the individual slopes and intercepts across 1000 iterations of resampling. 

\begin{figure*}
    \centering
    \includegraphics[width=0.8\linewidth]{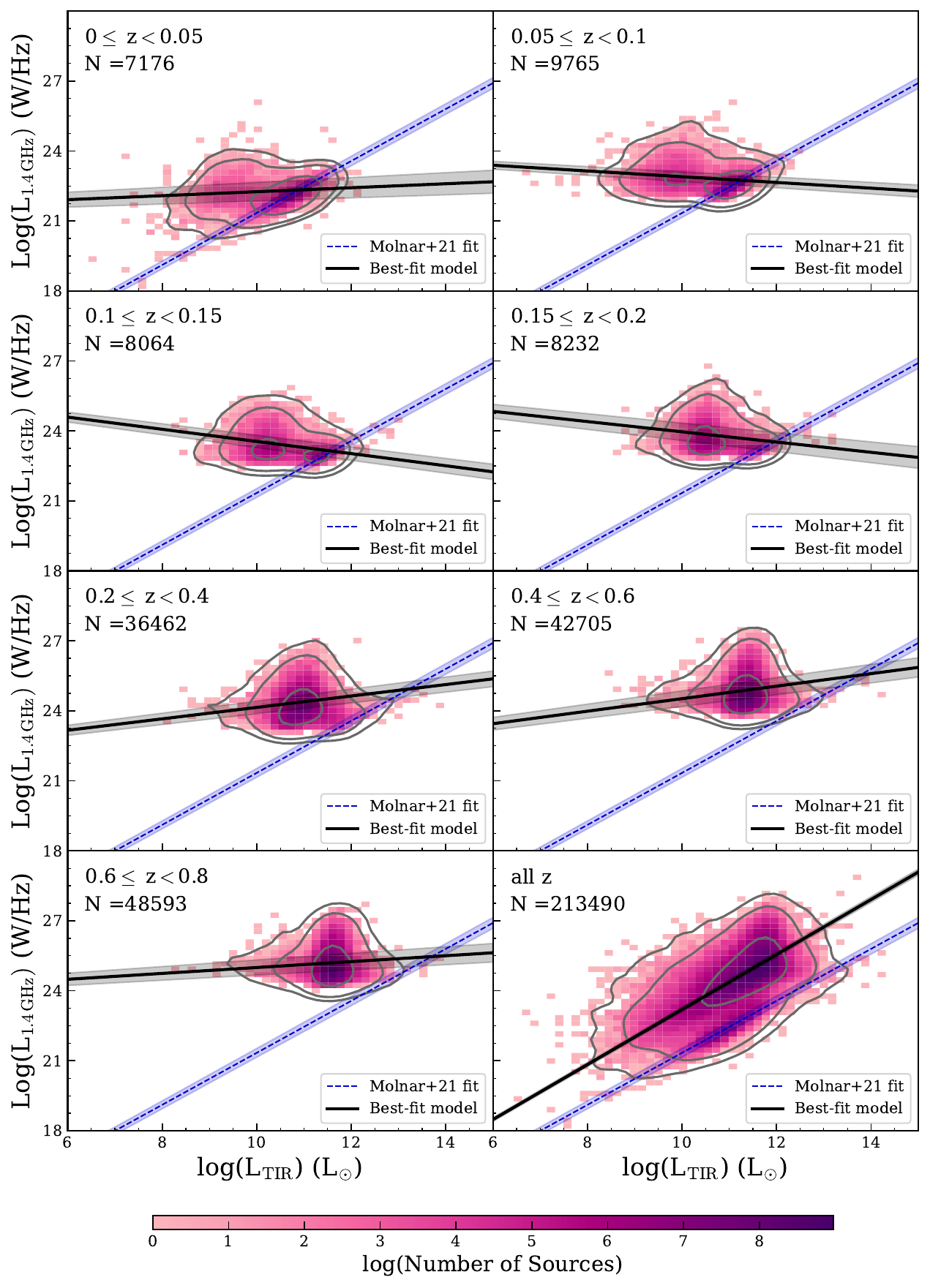}
    \caption{The 1.4 GHz vs. TIR luminosities for the model-selected galaxy sample across different redshift bins. The colour map indicates the logarithm of the number of sources in each pixel and the contours show the 10th, 50th, and 90th percentiles. For each graph, the redshift range and the number of sources is specified in the top left corner. The best-fit linear trend for each graph is shown in black, while the dashed blue line shows the relationship from Table 5 of \citet{Molnar2021}. For both, the shaded region indicates the associated $1\sigma$ confidence interval. The bottom-right panel shows the results for the full dataset.}
    \label{fig:lradio_ltir_subplots}
\end{figure*}

Figure \ref{fig:lradio_ltir_subplots} shows that the IRRC for the model-selected galaxy sample is not consistent with that of \citet{Molnar2021} across any redshift range. When all data is included, as in the bottom-right panel of Figure \ref{fig:lradio_ltir_subplots}, the slope of the IRRC is consistent with that of \citet{Molnar2021}, but there is a vertical shift of $\sim 1.2$. However, at lower redshifts, the fits are vastly different to \citet{Molnar2021}. For $z<0.1$, the fits deviate due to the presence of a subsample of galaxies with higher radio luminosities and lower TIR luminosities, which flatten the slope of the best-fit model. As $z$ increases, the dataset suffers from Malmquist bias, with a dearth of sources with $\rm log(L_{1.4 \, GHz}) \lesssim22$, indicating the sample is incomplete beyond $z\sim0.05-0.1$ due to the shallowness of RACS-mid. This is consistent with Figure \ref{fig:lradio_ltir_z}, which shows that these high-luminosity sources also have higher mean redshifts.

To account for these effects, the RACS-mid and WISE catalogues were depth-matched as in \citet{Molnar2021}, using the process described in Section \ref{sec:depth_matching} below. In addition, the higher-completeness, low-$z$ population is considered separately for the calculation of the cosmic SFRD in Section \ref{sec:sfrd}.

\begin{figure}
    \centering
    \includegraphics[width=0.99\linewidth]{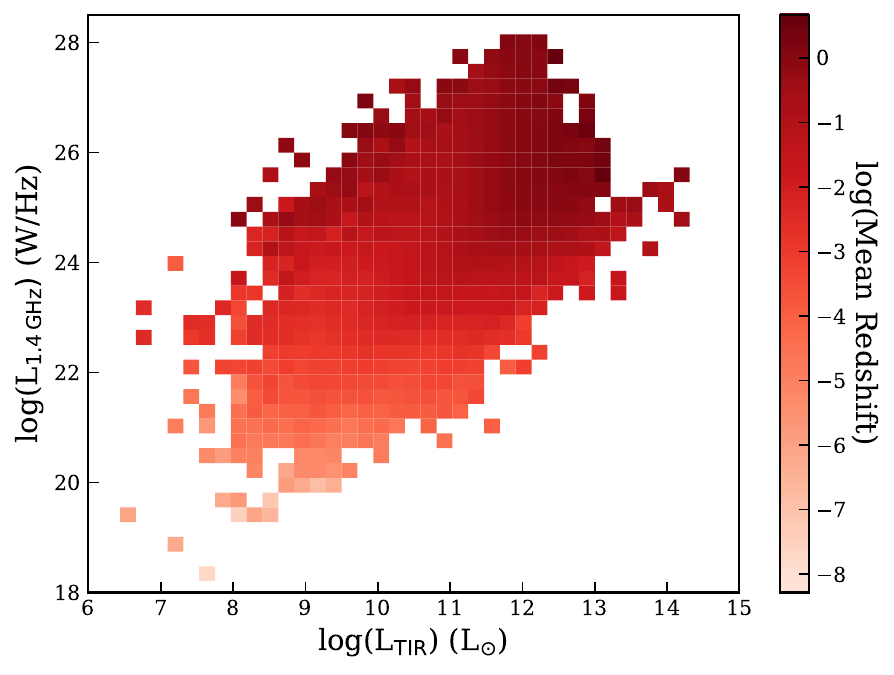}
    \caption{Similar to Figure \ref{fig:lradio_ltir_subplots} but with a colour map showing the logarithm of the mean redshift in each pixel.}
    \label{fig:lradio_ltir_z}
\end{figure}

\subsection{Depth-Matching the Radio and Infrared Catalogues}\label{sec:depth_matching}

First, to reduce noise in the dataset, an initial signal-to-noise cut was made on the model-selected galaxy sample, removing any sources with S/N$<5\sigma$. This reduced the sample size to 213,490 sources. 

Then, to correct for the different sensitivities of the WISE and RACS-mid catalogues, the model-selected galaxy sample was cleaned to ensure both surveys probe equivalent depths, following \citet{Sargent2010} and \citet{Molnar2021}. To do this, the luminosity corresponding to the $5\sigma$ flux-density limit of the RACS-mid catalogue (1.6 mJy, \citealt{Duchesne2024}) was calculated using equation \ref{eq:lum} at the redshift associated with each galaxy. 

Next, to convert the 1.4 GHz radio luminosities to their corresponding TIR luminosities, the median $q$ parameter of 2.54 from \citet{Molnar2021} was used. The parameter $q$ is often used to characterise the IRRC \citep{Helou1985}, and relates the TIR and radio luminosities to each other via:

\begin{equation} \label{eq:q}
    \rm q_{TIR} = log \bigg( \frac{L_{TIR}}{3.75 \times 10^{12} \, W}\bigg) - log \bigg( \frac{L_{1.4 \, GHz}}{W\, Hz^{-1}}\bigg) .
\end{equation}

Finally, equations \ref{eq:lum_w3pah} and \ref{eq:lum} were inverted to determine the W3PAH flux density, which corresponds to the flux limit of the RACS-mid catalogue. Any sources in the model-selected galaxy sample following the S/N cut with W3PAH flux densities below this value were removed.

This process removed 200,165 of 213,490 sources in the model-selected galaxy sample following the S/N cut. 5,648 of these excluded sources had $z<0.1$, and a further 14,410 had $0.1<z<0.2$. This left 11,293 ($\sim67\%$) and 1,886 ($\sim12\%$) of the original sources in each of these redshift ranges, respectively. Note that due to the shallowness of RACS-mid, nearly all sources with $z\gtrsim0.2$ were excluded due to falling below the 1.4 GHz sensitivity limit. 

Given that WISE is a much deeper survey, the RACS flux-density threshold was expected to limit the depth of the model-selected galaxy sample. To verify this, the same process as described above was performed in reverse using the flux-density limits for the WISE W1 and W3 bands (0.08 and 1 mJy, respectively; \citealt{Wright2010}). However, all model-selected galaxies had fluxes above this threshold, and thus no sources were excluded based on this criterion. The different samples referred to in this paper at each stage of the classification and cleaning process are summarised in Table \ref{tab:sample_summary}. 

\begin{table*}[]
\centering
\caption{Description and sizes of the different samples referred to in this study at each stage of the classification and cleaning process.}
\begin{tabular*}{\textwidth}{@{\extracolsep{\fill}} l l l}
\toprule \toprule
Sample Name & Description & Size \\
\toprule
Initial Cross-match & The sample obtained by cross-matching &  426,815 ($\sim$78\% galaxies, $\sim$13\% quasars, \vspace{-0.1cm} \\
Sample & WISE-PS1-STRM with RACS-mid. &  $\sim6$\% unsure, $\sim$3\% stars)\\
Final Cross-match & The initial cross-match sample after removing sources & 389,392 ($\sim$86\% galaxies, $\sim$14\% quasars) \vspace{-0.1cm} \\ 
Sample  & listed as stars or unsure in WISE-PS1-STRM. & \\
Model-selected Galaxy & The sources from the cross-match sample after star & 336,674 \vspace{-0.1cm} \\
Sample & removal classified as galaxies by the model. & \\
Model-selected Quasar  & The sources from the cross-match sample after star & 52,718 \vspace{-0.1cm} \\
 Sample & removal classified as quasars by the model. & \\
Model-selected Galaxy & Sources in the galaxy sample with a S/N $>5$. & 213,490 \vspace{-0.1cm} \\
Sample (S/N Cut) &  &  \\
Depth-matched Galaxy  & The remaining galaxies from the galaxy sample & 13,325 \vspace{-0.1cm} \\
 Sample & (S/N cut) after being corrected for the different & \vspace{-0.1cm} \\
 & sensitivities of the WISE and RACS-mid surveys. & \\
 $z<0.1$ Depth-matched & The low-redshift depth-matched galaxy sample. & 11,293 \vspace{-0.1cm} \\
 Galaxy Sample &  & \\
\toprule
\end{tabular*}
\label{tab:sample_summary}
\end{table*}

The 1.4 GHz and TIR luminosities of the depth-matched galaxy sample are shown in Figure \ref{fig:lradio_ltir_cleaned}. The top, middle, and bottom panels show the data for $z<0.1$, $0.1<z<0.2$, and $z<0.2$, respectively. As in \ref{fig:lradio_ltir_subplots}, the colour map indicates the logarithm of the number of sources in each pixel, and the IRRC from \cite{Molnar2021} is shown as the dashed blue line. Evidently, the depth-matching process removed the subsample of galaxies with higher radio and lower TIR luminosities that were flattening the slope of the IRRC at redshifts less than $\sim0.1$. As a result, the best-fit linear models, obtained using the bootstrapping technique described in Section \ref{sec:lums} above, are consistent with that of \citet{Molnar2021} ($m=1.114 \pm 0.009$, $c=10.2 \pm0.1$) and also \citet{Bell2003} for $z\lesssim0.2$. For $z<0.1$, $0.1<z<0.2$, and $z<0.2$, the best-fit parameters are $m=0.84\pm0.03$ and $c=13.24\pm0.31$, $m=0.92\pm0.06$ and $c=12.60\pm0.63$, and $m=0.94\pm0.02$ and $c=12.17\pm0.27$, respectively. Since $m\ne1$ for any of these fits, there is evidence for a slight degree of non-linearity in the $z<0.1$ depth-matched galaxy sample, in agreement with the results of \citet{Yun2001}, \citet{Bell2003}, and \citet{Molnar2021}. 

Another important implication of Figure \ref{fig:lradio_ltir_cleaned} is that there is evidence for sample incompleteness beyond $z\sim0.1$. This is indicated by the scarcity of sources with $\rm log(L_{TIR}) \lesssim 11$ and $\rm log(L_{1.4 \, GHz}) \lesssim 22.5$ in the middle panel, which shows $0.1<z<0.2$. As such, the best-fit models to the data for $z\gtrsim1$ are unlikely to constrain the true behaviour of the IRRC, as the lowest luminosity members of the population are not accounted for beyond this redshift. Thus, in the analysis in Section \ref{sec:sfrd}, the 1.4-GHz SFR calibration is modified according to the parameters of the IRRC for $z<0.1$, since the depth-matched galaxy sample is relatively complete at these redshifts, while for $z>0.1$, the calibration from \citet{Molnar2021} is used instead. 

\begin{figure}
    \centering
    \includegraphics[width=0.9\linewidth]{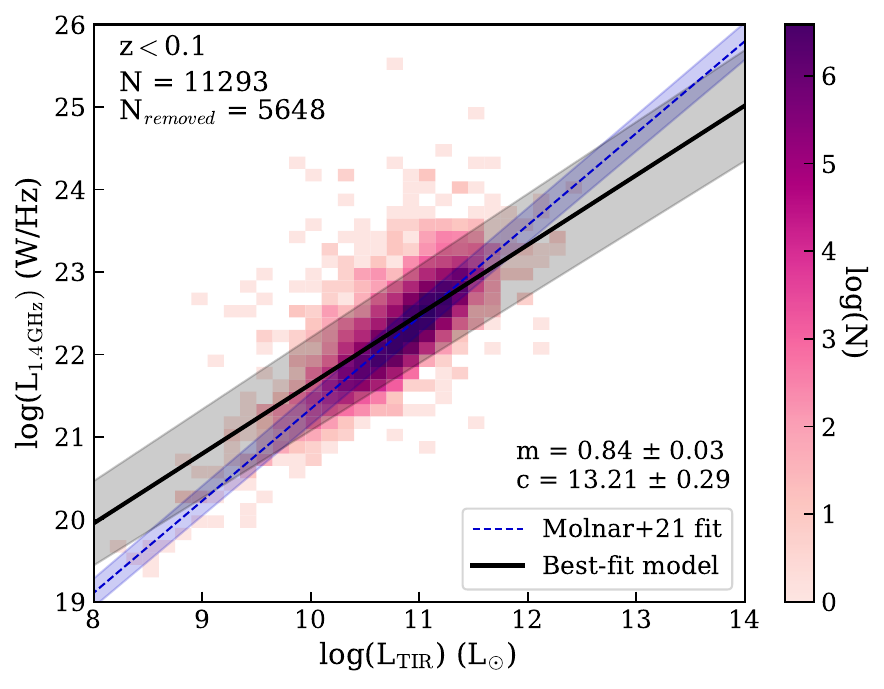} \\
    \includegraphics[width=0.9\linewidth]{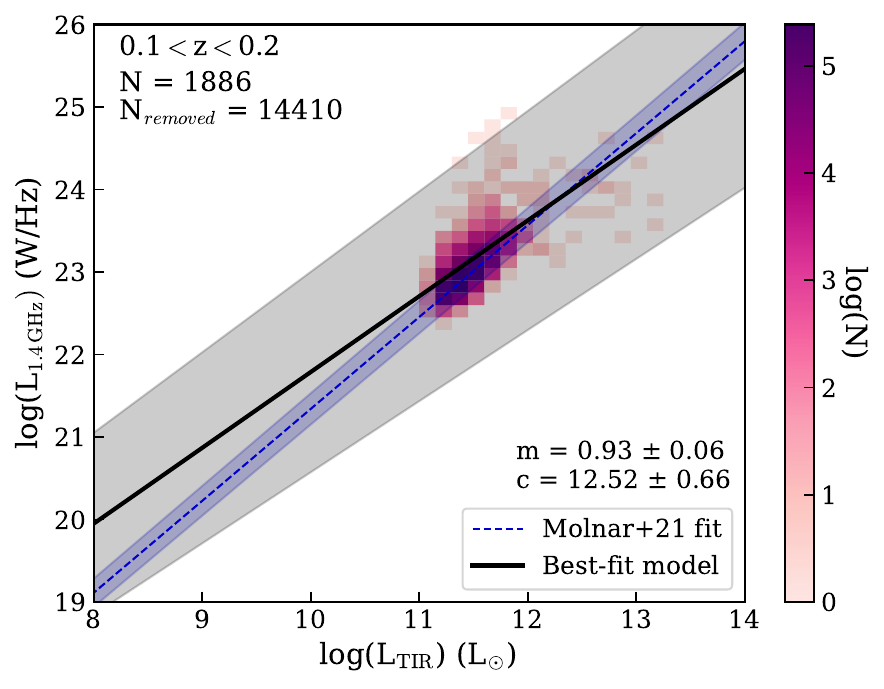}
    \includegraphics[width=0.9\linewidth]{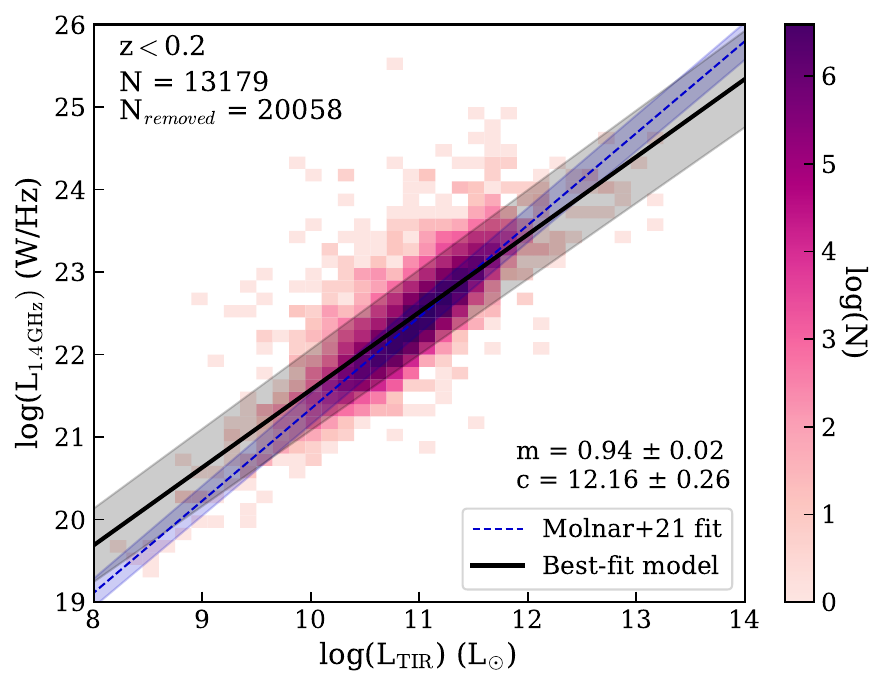}
    \caption{Similar to Figure \ref{fig:lradio_ltir_subplots} but for the depth-matched galaxy sample at $z<0.1$ (top), $0.1<z<0.2$ (middle), and for $z<0.2$ (bottom). The parameters of the best-fit linear model are shown bottom right. For each graph, the number of sources ($N$) and the number of sources removed through the depth-matching process ($N_{removed}$) are displayed top left.}
    \label{fig:lradio_ltir_cleaned}
\end{figure}

Figure \ref{fig:lr_dists} shows the 1.4 GHz radio and the TIR luminosity distributions calculated for the model-selected and depth-matched galaxy samples. Compared to the luminosity distributions obtained by \citet{Yun2001}, \citet{Bell2003}, and \citet{Molnar2021}, the model-selected galaxy sample is brighter in both radio continuum and IR. However, these studies considered only local galaxies ($z\lesssim0.1-0.2$) to maximise sample completeness. Using the depth-matched galaxy sample with $z<0.2$, the 1.4 GHz and TIR luminosity distributions become consistent with those of \citet{Molnar2021}, similar to those of \citet{Yun2001}, and overlap with those of \citet{Bell2003}. Therefore, the skew of the distributions towards high luminosities for the full population is likely due to a selection effect, since only the brightest sources are detected at extreme redshifts. In addition, the large redshift uncertainties for the high-$z$ sources likely also contributes to this effect, with $\sim10\%$ of the $z>0.1$ population having a relative uncertainty of $25\%$ or more.

\begin{figure*}
    \centering
    \includegraphics[width=0.48\linewidth]{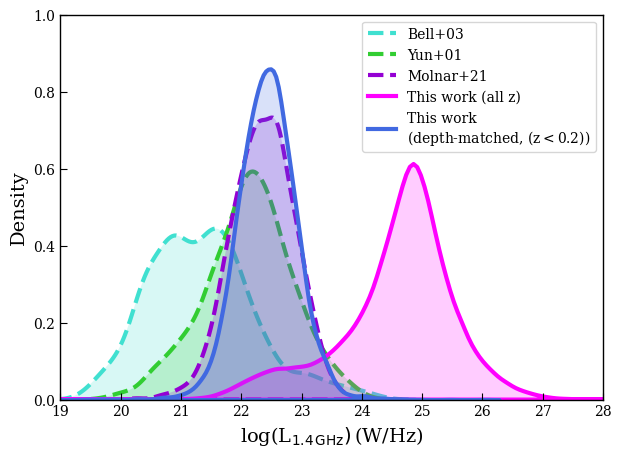}
    \includegraphics[width=0.48\linewidth]{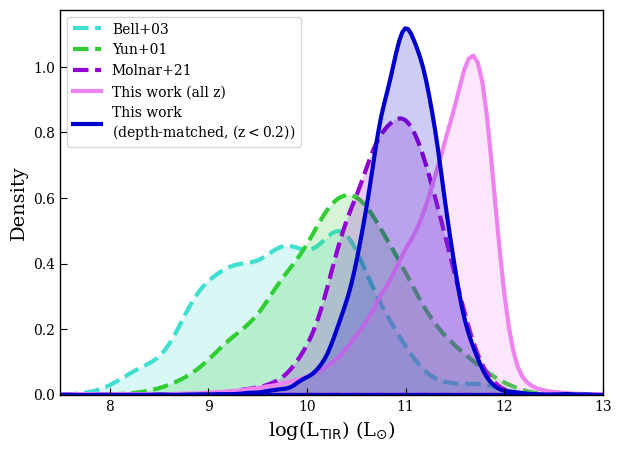}
    \caption{The 1.4 GHz radio (left) and TIR (right) luminosity distributions for those objects selected as galaxies by the model. The distributions are compared to those from \citet{Bell2003} (turquoise), \citet{Yun2001} (green), and \citet{Molnar2021} (purple). In each graph, the pink line shows the distribution of all model-selected galaxies, while the blue line is the distribution for the depth-matched galaxy sample with $z<0.2$. Note that these distributions are kernel density estimates, and therefore the probability is given by the area under the curve, which sums to one.}
    \label{fig:lr_dists}
\end{figure*}

\subsection{Using the IRRC to estimate the SFRs of the $z<0.1$ population}\label{sec:irrc}

It is currently well established that the 1.4 GHz emission from SFGs is correlated with their SFR \citep{Condon1992, Murphy2011a, Murphy2012, Tabatabaei2017, Gurkan2018, Smith2021}. However, such studies often rely on SFR prescriptions that do not allow for or consider the case of a non-linear IRRC, for which there is increasing evidence \citep[e.g.,][]{Yun2001, Bell2003, Molnar2021}. Therefore, \citet{Molnar2021} developed a new 1.4-GHz SFR calibration which accounts for this non-linear behaviour and can be modified based on the parameters of the IRRC. In this section, we determine a modified 1.4-GHz SFR prescription from the best-fit parameters to the IRRC that is used to calculate the SFRD of the model-selected SFGs in Section \ref{sec:sfrd}.

A graph of $q_{TIR}$ vs. $\rm log(L_{1.4 \, GHz})$ (see equation \ref{eq:q}) should follow a linear relationship with slope $m$ and intercept $c$ of the form $q_{TIR} = m \rm \, log(L_{1.4\,GHz}) + c$. The 1.4-GHz SFR calibration is then modified according to the best-fit slope and intercept using Equation 21 from \citet{Molnar2021}:

\begin{equation} \label{eq:SFR_general}
    \rm log(SFR) = (\textit{m} + 1) \, log(L_{1.4 \, GHz}) + (\textit{c} + \textit{f}_{\textit{scale}}),
\end{equation} 

\noindent{where $f_{scale}= 10^{-10} \rm M_{\odot}\,yr^{-1}\,L_{\odot}$ is the $\rm L_{TIR}$-SFR scaling factor used in \citet{Molnar2021} based on the work of \citet{Kennicutt1998} and \citet{Madau2014}. 

In their study and using a depth-matched sample of $z<0.2$ SFGs, \citet{Molnar2021} obtained a best-fit slope and intercept of $m=-0.177 \pm 0.009$ and $c=6.5 \pm 0.2$, respectively. Substituting these values into Equation \ref{eq:SFR_general} gave the following 1.4 GHz SFR calibration \citep[Equation 22,][]{Molnar2021}:

\begin{multline} \label{eq:sfr_molnar}
    \rm log \bigg( \frac{SFR}{M_{\odot} \, yr^{-1}}\bigg) = (0.823 \pm 0.009) \, log \bigg( \frac{L_{1.4 \, GHz}}{W \, Hz^{-1}} \bigg) \\ - (17.5 \pm 0.2).
\end{multline}

Figure \ref{fig:q_value_plots_num} shows $q_{TIR}$ versus $\rm log(L_{1.4 \, GHz})$ for the depth-matched galaxy sample. The black line shows the best-fit linear model for $z<0.1$, while the dotted red line shows the fit for $z<0.2$. These fits are compared to that obtained by \citet{Molnar2021} in orange. The best-fit slopes and intercepts are given in Table \ref{tab:qtir_fit_params}. Both fits agree within $\sim1\sigma$ with that of \citet{Molnar2021}. However, since the completeness of the RACS-mid catalogue is degraded beyond $z\sim0.1$, the fit for $z<0.1$ was used to obtain a modified SFR prescription for these redshifts (see Section \ref{sec:sfrd}), while for $z>0.1$, the relation from \citet{Molnar2021} in Equation \ref{eq:sfr_molnar} was used instead, since their sample was complete out to a redshift of $\sim0.2$. 

\begin{table}[]
\centering
\caption{Best-fit linear model parameters to $q_{TIR}$ versus $\rm log(L_{1.4 \, GHz})$ (Figure \ref{fig:q_value_plots_num}).}
\begin{tabular*}{0.48\textwidth}{@{\extracolsep{\fill}} l c c}
\toprule \toprule
Sample & $m$ & $c$ \\
\toprule
 Depth-matched $z<0.1$ & $-0.255 \pm 0.088$ & $8.2 \pm 1.9$ \\
 Depth-matched $z<0.2$ & $-0.199\pm0.044$ & $7.0 \pm 1.0$  \\
 \citet{Molnar2021} & $-0.177 \pm 0.009$ & $6.5 \pm 0.2$\\
\toprule
\end{tabular*}
\label{tab:qtir_fit_params}
\end{table}

\begin{figure}
    \centering
    \includegraphics[width=0.99\linewidth]{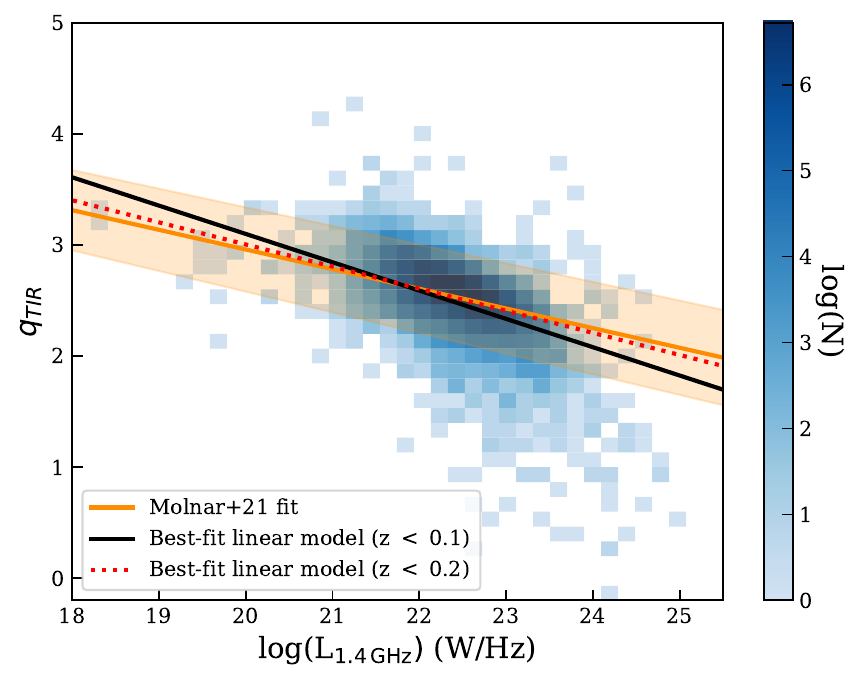}
    \caption{Graph of $q_{TIR}$ vs $\rm log(L_{1.4 \, GHz})$ for the depth-matched galaxy sample with $z<0.1$. The colour map corresponds to the logarithm of the number of sources in each pixel. The black line is the best-fit linear model to the $z<0.1$ data, and the dotted red line is the fit for $z<0.2$. The uncertainties in these models are not shown as they cover the majority of the graph. The orange line is the fit from \citet{Molnar2021}.}
    \label{fig:q_value_plots_num}
\end{figure}

To tailor the 1.4-GHz SFR prescription from \citet{Molnar2021} for use with the $z<0.1$ depth-matched galaxy sample, Equation \ref{eq:SFR_general} was modified using the best-fit parameters from Figure \ref{fig:q_value_plots_num}, giving:

\begin{multline} \label{eq:SFR_specific}
    \rm log(SFR_{low\, z}) = (0.745 \pm 0.088) \, log(L_{1.4 \, GHz}) \\+ (-15.8 \pm 1.9).
\end{multline} 

\section{Completeness Correction}\label{sec:completeness_corr}

Since Figures \ref{fig:lradio_ltir_subplots} and \ref{fig:lradio_ltir_cleaned} indicate that the model-selected galaxy sample is incomplete for $z\gtrsim0.1$, some form of completeness correction is required to approximate the true SFRD. Figure \ref{fig:luminosity_function} shows the luminosity function of the $z<0.1$ depth-matched galaxy sample, which was calculated by binning the sample according to luminosity and weighting each galaxy's contribution by the inverse of the maximum comoving volume in which it could be observed given the RACS-mid flux limit. A flux limit of 1.6 mJy for RACS-mid was adopted, at which level 95\% completeness is expected \citep{Duchesne2024}. All 11,293 sources in the $z<0.1$ depth-matched galaxy sample were used to construct the luminosity function presented in Figure \ref{fig:luminosity_function}.

\begin{figure}
    \centering
    \includegraphics[width=0.99\linewidth]{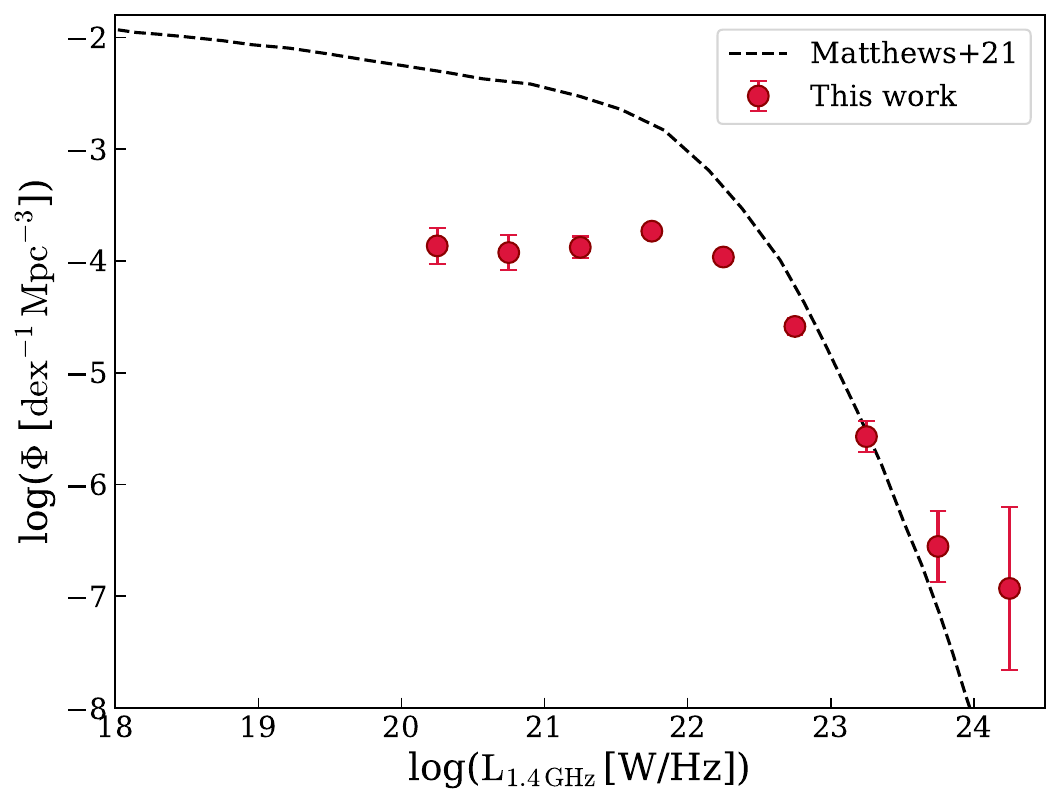}
    \caption{The luminosity function $\Phi$ (red data points) of the model-selected galaxies with $z<0.1$, compared to the local luminosity function (dashed) from Figure 1 of \citet{Matthews2021}.}
    \label{fig:luminosity_function}
\end{figure}

Figure \ref{fig:luminosity_function} also shows the luminosity function derived from the energy-density function in Figure 1 of \citet{Matthews2021}. \citet{Matthews2021} obtained this results using a spectroscopically complete sample of $\sim10^4$ NVSS galaxies. As such, it serves as a baseline from which it is possible to estimate the completeness of the $z<0.1$ depth-matched galaxy sample. To do this, the luminosity functions in Figure \ref{fig:luminosity_function} were multiplied by luminosity and integrated using \texttt{scipy.integrate.simpson} between $18 < \rm log(L_{1.4 \, GHz}) \leq 22.5$, since $\rm log(L_{1.4 \, GHz}) \sim 22.5$ is the approximate luminosity below which the completeness of the sample is degraded. This is indicated by the divergence at this luminosity of the model-selected galaxy sample and \citet{Matthews2021} luminosity functions in Figure \ref{fig:luminosity_function}, as well as the windowing effect that is apparent below this limit in \ref{fig:lradio_ltir_subplots} and \ref{fig:lradio_ltir_cleaned}. Mathematically, the luminosity completeness may be expressed as:

\begin{equation}
    \rm completeness = \it \frac{\int L_1\,\phi_1(L) \,dL_1}{\int L_2\,\phi_2(L) \,dL_2}
\end{equation}

\noindent where $L$ is the 1.4 GHz luminosity, $\phi(L)$ is the luminosity function, and the subscripts 1 and 2 denote the $z<0.1$ depth-matched galaxy and \citet{Matthews2021} samples, respectively. 

Taking the ratio of these integrals gives the completeness of the luminosity contribution of the $z<0.1$ depth-matched galaxy sample, which evaluates to $C_{z<0.1} = 0.11$. Therefore, dividing the local SFRD determined in Section \ref{sec:sfrd} by this value will yield an approximation of the true value. However, it is important to note that large completeness corrections, which are common in SFRD studies, have the effect of amplifying small errors in the SFR. Therefore, using deeper, higher completeness surveys is key to reducing these uncertainties and improving the robustness of future SFRD estimates.

\section{Discussion}\label{sec:sfrd}

For the 11,293 sources in the $z<0.1$ depth-matched galaxy sample, the modified prescription in Equation \ref{eq:SFR_specific} was used to calculate their SFRs. In addition, the SFRs of all 13,325 sources in the full depth-matched galaxy sample including those with redshifts greater than 0.1 were calculated using the \citet{Molnar2021} prescription in Equation \ref{eq:sfr_molnar}. Using this data and the photometric redshifts from \citet{Beck2022}, the SFRD as a function of redshift can be approximated. 

To do this, galaxies are first binned according to redshift. For the $z<0.1$ depth-matched sample, a single redshift bin from $(0, 0.1)$ was used, while for the full depth-matched sample, similar bins to those in \citet{Matthews2024} were used, i.e. $(0,0.2), (0.2, 0.4)$. The higher redshift bins from \citet{Matthews2024} were not included given the very low completeness beyond $z\sim0.1$ due to the S/N cuts and depth-matching procedure described in Section \ref{sec:depth_matching}, with only 115 sources with redshifts between 0.2 and 0.4. As a result of this incompleteness, the SFRD will be significantly underestimated beyond $z\sim0.1$. The total SFR in a given epoch is then obtained by summing the SFRs of all galaxies in each bin, which are calculated using either the $z<0.1$ prescription from the fit to the IRRC (Equation \ref{eq:SFR_specific}), or the prescription from \citet{Molnar2021} (Equation \ref{eq:sfr_molnar}). 

As for the luminosity function in Section \ref{sec:completeness_corr}, the observational bias associated with the flux-limited RACS-mid sample is accounted for using the $1/V_{max}$ method from \citet{Schmidt1968}. Specifically, fainter objects are detectable within smaller comoving volumes. Therefore, by weighting each galaxy's contribution to the total SFR by the inverse of the maximum comoving volume in which it could be observed given a flux limit of 1.6 mJy \citep{Duchesne2024}, the underlying distribution of the cosmic SFRD is obtained. The SFRD values were also corrected for the sky coverage of the cross-matched RACS-mid and WISE catalogues ($\sim64\%$). The SFRD results and uncertainties are summarised in Table \ref{tab:sfrd_results}.

\begin{table*}[]
\centering
\caption{Summary of the SFRD results obtained in this work that are shown in Figure \ref{fig:sfr_plots}. $\sigma_{SFR}$ is uncertainty associated with the SFR of each galaxy, which includes the uncertainty in the 1.4 GHz luminosity as well as the uncertainty due to the linear fit used to determine the $z<0.1$ SFR prescription in Equation \ref{eq:SFR_specific} or the fit from \citet{Molnar2021} in Equation \ref{eq:sfr_molnar}. $\sigma_{total}$ additionally includes the uncertainty associated with $V_{max}$, propagated in quadrature.}
\begin{tabular*}{0.99\textwidth}{@{\extracolsep{\fill}} l c c c c c}
\toprule \toprule
Redshift Bin & Number of Sources & \makecell{\rule{0pt}{2ex} SFRD \\ ($\rm M_{\odot}\,yr^{-1}\,Mpc^{-3}$)} & \makecell{\rule{0pt}{2ex} $\sigma_{SFR}$ \\ ($\rm M_{\odot}\,yr^{-1}\,Mpc^{-3}$)} & \makecell{\rule{0pt}{2ex}$\sigma_{total}$ \\ ($\rm M_{\odot}\,yr^{-1}\,Mpc^{-3}$)} & SFR Prescription  \\
\toprule \toprule
0 -- 0.1 & 11,293 & $1.6 \times 10^{-3}$ & $5.4 \times 10^{-5}$ & $5.5 \times 10^{-4}$ & Eq. \ref{eq:SFR_specific}  \\
0 -- 0.1 & 11,293 & $1.4 \times 10^{-2} $ & $4.7\times 10^{-4}$ & $4.8\times 10^{-3}$ & Eq. \ref{eq:SFR_specific} (corrected)\\
0 -- 0.2 & 13,179 & $1.4 \times 10^{-3} $ & $3.8 \times 10^{-5} $ & $4.5 \times 10^{-5}$ & Eq. \ref{eq:sfr_molnar} \\
0.2 -- 0.4 & 115 & $1.2 \times 10^{-5}$ & $6.1 \times 10^{-6} $ & $6.4 \times 10^{-6} $& Eq. \ref{eq:sfr_molnar} \\
\toprule
\end{tabular*}
\label{tab:sfrd_results}
\end{table*}

Figure \ref{fig:sfr_plots} shows the evolution of the SFRD as a function of redshift that was obtained using the $z<0.1$ depth-matched galaxy sample and the modified SFR prescription (Eq. \ref{eq:SFR_specific}, and the \citet{Molnar2021} prescription and the full depth-matched galaxy sample. The completeness corrected $z<0.1$ results are also shown (see Section \ref{sec:completeness_corr}). These results are compared to previously published SFRDs using radio \citep{Mauch2007, Karim2011, Upjohn2019, Brown2020, Enia2022, Malefahlo2022, Cochrane2023, Matthews2024}, UV \citep{Cucciati2012, Madau2014, Bouwens2015}, IR \citep{Gruppioni2013}, and multi-wavelength \citep{Hopkins2006, Behroozi2013, Dunlop2017, Driver2018} observations. Each of the SFRDs calculated as part of this paper are shown with two sets of error bars. The coloured error bars are the uncertainty associated with the SFR of each galaxy, which includes the uncertainty in the 1.4 GHz luminosity as well as the uncertainty due to the linear fit used to determine the $z<0.1$ SFR prescription in Equation \ref{eq:SFR_specific} or the fit from \citet{Molnar2021} in Equation \ref{eq:sfr_molnar}. The grey error bars additionally include the uncertainty associated with $V_{max}$, propagated in quadrature. The uncertainty in $V_{max}$ is tied to the photometric redshift errors provided in \citet{Beck2022}, for which a small fraction of the total sample (0.5\%) exceed 100\% relative uncertainty. In addition, the error in $V_{max}$ is large for sources that are below or near the survey detection limit of 1.6 mJy. Since this is more common for distant sources, this results in the very large grey error bars seen at higher redshifts. The total uncertainties in SFRD using the \citet{Molnar2021} calibration are so large that for $\sim80\%$ of the data points, they extend beyond the range of the graph. Future spectroscopic surveys such as the 4MOST Hemisphere Survey (4HS, \citealt{Taylor2023}) will allow more precise redshifts to be determined, and deeper radio surveys such as the Evolutionary Map of the Universe project (EMU, \citealt{Norris2011, Norris2021}) will reduce the number of sources that are detected near the survey flux limit. Together, these effects will act to reduce the uncertainty associated with the SFRD calculation in future studies. Notably, the error bars associated with the $z<0.1$ result are larger than the error in the lowest redshift bin of the \citet{Molnar2021} prescription result. This is because the uncertainties in the parameters of the modified $z<0.1$ SFR prescription in Equation \ref{eq:SFR_specific} are $\sim10$ times larger than that of Equation \ref{eq:sfr_molnar}. 

In addition to the uncertainty associated with the SFR and the volumetric correction, further uncertainty is introduced during the machine-learning classification process itself. For example, the optimised \texttt{XGBClassifier} model achieved an accuracy of $\sim94\%$, meaning that $\sim6\%$ of sources have incorrect predictions. Since the model in \citet{Beck2022} only predicted photometric redshifts for sources they classified as galaxies, no quasars incorrectly classified as galaxies by the optimised model were included in our SFRD calculation, nor were any sources classified as galaxies that did not have photometric redshifts in the \citet{Beck2022} catalogue. However, galaxies that were missed by the model and incorrectly classified as quasars were not included in the final result. This would have caused the SFRD to be somewhat underestimated prior to being corrected for sample incompleteness. This effect is likely to be more significant at higher redshifts ($z\gtrsim0.5$), where the model more frequently struggled to classify sources (see Section \ref{sec:redshift_effects}).

\begin{figure*}
    \centering
    \includegraphics[width=0.99\linewidth]{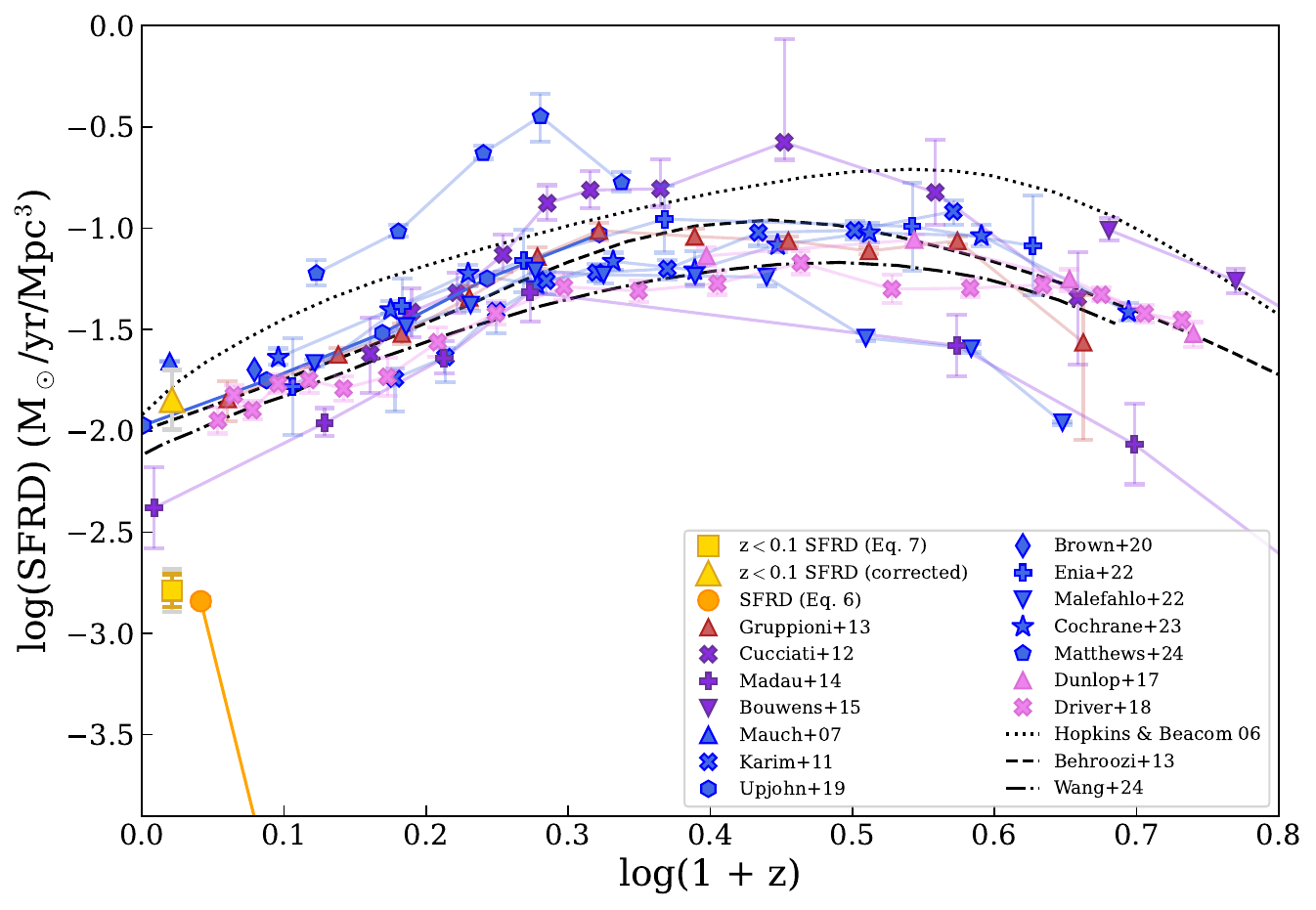}
    \caption{The cosmic SFRD as a function of $\rm log(1+ \textit{z})$ using the $z<0.1$ prescription (yellow square) and the \citet{Molnar2021} prescription (orange circles). The yellow triangle shows the $z<0.1$ value after being corrected for sample incompleteness. The coloured error bars are those derived using the SFR uncertainties, while the grey error bars also include the uncertainty in the maximum comoving volume for each galaxy. These results and the data sets from which they were calculated are also summarised in Table \ref{tab:sfrd_results}. For comparison, also included are the SFRDs obtained using radio in blue \citep{Mauch2007, Karim2011, Upjohn2019, Brown2020, Enia2022, Malefahlo2022, Cochrane2023, Matthews2024}, UV in purple \citep{Cucciati2012, Madau2014, Bouwens2015}, IR in red \citep{Gruppioni2013}, and multi-wavelength observations in pink \citep{Dunlop2017, Driver2018}. The analytical fits of \citet{Hopkins2006}, \citet{Behroozi2013} and \citet{Wang2024} are also included for comparison. Note that due to the low completeness beyond $z\sim0.1$, the series in orange drops off sharply and the highest redshift data point is not shown. The exact value can be found in Table \ref{tab:sfrd_results}.}
    \label{fig:sfr_plots}
\end{figure*}

Furthermore, the accuracy of the model was assessed against the labels from \citet{Beck2022}, which were also the output of a machine-learning model, specifically a dense neural network. Their model achieved precision scores on the blind test set of 99.34\% for galaxies and 97.82\% for quasars, implying that up to $\sim1-2$\% of labels that were used as truth values to train the optimised model may have been incorrect. Neither this nor the uncertainty associated with the optimised model's predictions are accounted for in the error bars in Figure \ref{fig:sfr_plots}. However, due to the large area covered by the overlapping regions of RACS-mid and WISE ($\sim64\%$), the values are not expected to be affected by cosmic variance. 

Comparing the $z<0.1$ SFRD to previously published results, the uncorrected value of $(1.6 \pm 0.6) \times 10^{-3} \; \rm M_{\odot}\,yr^{-1}\,Mpc^{-3}$ obtained in this paper for $z < 0.1$ is significantly lower than that obtained in other studies, including \citet{Mauch2007} and \citet{Upjohn2019}, and the analytical fits of \citet{Hopkins2006} and \citet{Behroozi2013}, which were derived using published, multi-wavelength datasets. However, after correcting for sample incompleteness, the $z<0.1$ value ($(1.4 \pm 0.5) \times 10^{-2} \; \rm M_{\odot}\,yr^{-1}\,Mpc^{-3}$) is consistent with \citet{Mauch2007}, \citet{Hopkins2006}, \citet{Behroozi2013}, \citet{Upjohn2019}, and \citet{Wang2024}. While the completeness correction places the $z<0.1$ SFRD within the range expected from other studies, a longer integration time will more accurately constrain the value. In addition, since the $z<0.1$ and completeness-corrected values are not greater than that determined by other groups, the depth-matched galaxy sample does not appear to be significantly affected by contamination from AGN, as this would likely yield a SFRD that is higher than expected from other studies. This is consistent with the results of the cross-match with the Milliquas catalogue \citep{Flesch2023} described in Section \ref{sec:results}, which indicates an AGN contamination fraction of $\sim3\%$ for $z<0.1$.

The $z<0.1$ measurement suggests that machine-learning models can be leveraged to identify relatively uncontaminated samples of galaxies that can be used to probe the cosmic SFRD. As such, this study presents a proof of concept that machine-learning-identified samples can achieve results that are consistent with studies that select populations of SFGs using other methods, for example using spectroscopy. In addition, the SFR prescription from \citet{Molnar2021}, which can be modified depending on the IR and radio properties of the available galaxy sample, has been shown to provide results consistent with previous studies of the SFRD, using separate data and a different redshift range to that in \citet{Molnar2021}.

Considering the SFRD obtained using the \citet{Molnar2021} prescription in Equation \ref{eq:sfr_molnar} and the full depth-matched galaxy sample (orange series, Figure \ref{fig:sfr_plots}), the values are lower than those obtained by other groups, although this is not surprising given that these results were not corrected for incompleteness. The lowest redshift data point differs by $\sim12\%$ from the uncorrected value obtained using our modified SFR prescription, however they are consistent within uncertainties. This indicates that our analysis is not sensitive to the choice of $q_{TIR}$ that was used to depth-match the sample in Section \ref{sec:depth_matching} (i.e. the value from \citealt{Molnar2021}). The sharp decrease in the SFRD beyond the lowest redshift data point reflects the stark decrease in galaxy completeness beyond $z\sim 0.25$, however, this is not unexpected given the shallowness of RACS due to its short integration time of $\sim15$ minutes per field. Future, deeper ASKAP surveys, such as EMU \citep{Norris2011, Norris2021}, for which the results of the pilot survey were released in 2021, will enable the SFRD to be probed at higher redshifts due to a significant increase in survey completeness.

One caveat is that the SFRD results in this section relied on the photometric redshifts from \citet{Beck2022}. Without these, no redshift data would have been available, prohibiting the determination of the SFRD as a function of redshift. However, this paper serves as a proof of concept for the method, which uses one model to select a population of SFGs using multi-wavelength, tabular data, and a second model to predict photometric redshifts, similar to the one implemented in \citet{Beck2022}. Since RACS is a shallow survey with low completeness beyond $z\sim0.1$, EMU will provide a wealth of data ($\sim 70 \times 10^6$ sources) to be used as inputs for future machine-learning models such as the one implemented in this paper. In addition, the final EMU catalogue will include optical and IR cross-identifications along with photometric redshifts for approximately half of the sources. This will remove the need for a second model to predict photometric redshifts, and make the classification model more simple to implement.

\section{Summary and Conclusions}\label{sec:conc}

In this paper, we tested the utility of a gradient-boosted decision tree model to classify extragalactic radio sources as either galaxies or quasars using published wide-field IR and radio catalogues. 105 features from the RACS-mid and WISE All-sky catalogues were used as model inputs, and the machine-learning derived classifications from \citet{Beck2022} were used as labels to assess model performance, resulting in 389,392 sources in the final cross-matched sample. Given the significant class imbalance of the sample ($\sim86\%$ galaxies and $\sim14\%$ quasars), various resampling techniques were trialled to improve performance. Ultimately, an \texttt{XGBClassifier} model with random oversampling of the quasar minority class was shown to achieve optimal performance on a trial dataset consisting of $\sim58,000$ sources. This model was then extended to the full dataset, which was split into train, validation, and test sets using a 70\%-15\%-15\% split. Following hyperparameter tuning, the model achieved a weighted F1 score of 0.93 and an accuracy of 0.94 on the test dataset, ultimately classifying 336,674 sources as galaxies and 52,718 sources as quasars. By cross-matching with the Milliquas catalogue \citep{Flesch2023}, an AGN contamination rate of 0.05 (0.03 for $z<0.1$) was determined for the model-selected galaxy sample.

The SFRs of those objects classified as galaxies by the optimised model were then determined using both the prescription from \citet{Molnar2021} and a modified prescription based on the fit to the IRRC of the $z<0.1$ sample. Since the \citet{Beck2022} catalogue contains photometric redshifts for those sources they classified as galaxies, the SFRD as a function of redshift was calculated for the $z<0.1$ depth-matched galaxy sample identified by the model. Using the new $z<0.1$ prescription, a completeness-corrected local SFRD of $(1.4 \pm 0.5) \times 10^{-2} \; \rm M_{\odot}\,yr^{-1}\,Mpc^{-3}$ was determined using 11,293 sources, consistent with that obtained by \citet{Mauch2007} and \citet{Upjohn2019}, and the analytical fits of \citet{Hopkins2006} and \citet{Behroozi2013}. The values obtained using the \citet{Molnar2021} calibration that were not corrected for completeness but are consistent with our uncorrected value at low redshifts. The \citet{Molnar2021} prescription SFRD drops off above $z\sim0.25$, indicative of poor survey completeness beyond this epoch.  

Future directions include the extension of the model to predict the nature of all sources in the RACS-mid catalogue. Currently, the model has generated predictions for sources with counterparts and hence labels as either galaxies or quasars in the \citet{Beck2022} catalogue, amounting to 389,392 sources. However, RACS-mid consists of $\sim 3.1 \times 10^6$ sources whose classifications as either a galaxy or a quasar can be predicted by the model following a cross-match with the WISE All-Sky Catalogue. If redshift information is available, these sources can be included in the local SFRD estimate, and if not, photometric redshifts can be predicted for the entire catalogue by implementing a separate model as in \citet{Beck2022}.

The performance of the model may also be improved by the inclusion of data at other wavelengths, for example using optical observations as in \citet{Beck2022}, or UV data as in \citet{Karsten2023}. Additional performance improvements may also be possible with the implementation of a more complex model, such as the dense neural network used by \citet{Beck2022}. 

Despite good agreement between the local SFRD and previously published measurements, the cosmic SFRDs that are presented in Section \ref{sec:sfrd} suffer from large uncertainties as well as an inability to probe the distribution at high redshifts. This is because any machine-learning-based approach will retain the uncertainties associated with the photometric redshifts on which it was trained. However, future, deeper photometric and spectroscopic surveys such as EMU and 4HS will reduce these uncertainties and allow the behaviour of the SFRD to be studied at earlier times. This is because spectroscopic redshifts are significantly more precise and fewer sources will be detected near or below the survey detection limit, reducing uncertainties, and sufficient survey completeness will be achieved out to higher redshifts. Thus, the release of data from these currently underway surveys will permit the implementation of a new machine-learning model with improved performance which can select large, pure samples of galaxies, ultimately allowing for a more precise and global determination of the cosmic SFRD. Since all current estimates of the cosmic SFRD rely on large completeness or dust-attenuation corrections for which the uncertainties are possibly underestimated, complete, machine-learning selected samples of galaxies from very large 1.4 GHz surveys are an exciting prospect for future studies.

In summary, the forthcoming wide-field 1.4-GHz radio surveys will provide:

\begin{enumerate}
    \item improved statistical coverage of both low- and high-redshift radio sources,
    \item reduced impact due to cosmic variance,
    \item reliable galaxy-quasar separation using a combination of machine-learning models trained on radio and IR photometry
    \item robust estimates of the relationship between radio emission and star formation across redshift. 
\end{enumerate}

However, future challenges to be resolved include reliable redshift estimates, or preferably measurements, and the detection of sub-$\mu$Jy sources in deep radio surveys. 

\section*{Acknowledgements}
We thank Suk Yee Yong for the helpful discussions and advice relating to resampling techniques. 
We thank the anonymous referee for their insightful and constructive comments that greatly improved the clarity and rigour of this work.
This scientific work uses data obtained from Inyarrimanha Ilgari Bundara/the Murchison Radio-astronomy Observatory. We acknowledge the Wajarri Yamaji People as the Traditional Owners and native title holders of the Observatory site. CSIRO’s ASKAP radio telescope is part of the Australia Telescope National Facility, which is funded by the Australian Government with support from the National Collaborative Research Infrastructure Strategy. ASKAP uses the resources of the Pawsey Supercomputing Research Centre. Establishment of ASKAP, Inyarrimanha Ilgari Bundara, the CSIRO Murchison Radio-astronomy Observatory, and the Pawsey Supercomputing Research Centre are initiatives of the Australian Government, with support from the Government of Western Australia and the Science and Industry Endowment Fund. This paper also includes
archived data obtained through the CSIRO ASKAP Science Data Archive, CASDA, as well as data from the NASA/IPAC Infrared Science Archive, which is operated by the Jet Propulsion Laboratory, California Institute of Technology, under contract with the National Aeronautics and Space Administration. The WISE-PS1-STRM was accessed via the Mikulski Archive for Space Telescopes (MAST) CasJobs via \href{https://archive.stsci.edu/doi/resolve/resolve.html?doi=10.17909/wf64-kq10}{doi:10.17909/wf64-kq10}.

\section*{Funding Statement}
SP was supported by the Australian Government Research Training Program Scholarship, the University of Melbourne School of Physics, and the Dr Albert Shimmins Fund.

\bibliographystyle{paper} 
\bibliography{ref.bib}

\begin{appendix}

\renewcommand{\thefigure}{A\arabic{figure}}
\setcounter{figure}{0}

\renewcommand{\thetable}{A\arabic{table}}
\setcounter{table}{0}

\section{Feature Engineering and Hyperparameter Optimisation}

\subsection{RACS-mid and WISE Features Used as Model Inputs}\label{sec:features}

The WISE and RACS-mid features used as model inputs are listed in Table \ref{tab:feature list}.

\renewcommand{\arraystretch}{1.25}
\begin{table}[]
\centering
\small
\caption{The base WISE and RACS-mid features used as model inputs. The notation `$?$' indicates that the feature is measured in each of the four WISE passbands. For example, $w?mpro$ refers to the four profile-fit photometry magnitudes $w1mpro$, $w2mpro$, $w4mpro$, and $w4mpro$.}

\begin{tabular*}{0.48\textwidth}{@{\extracolsep{\fill}} l l c c }
\toprule \toprule
Feature Name &  Description \\
\toprule \toprule
WISE & \\
\toprule
$w?mpro$ &  Profile-fit photometry magnitudes \\
$w?sigmpro$ &  Profile-fit photometry magnitude uncertainties \\
$w?rchi2$ &  Reduced chi-square of the profile-fit \vspace{-0.1cm} \\
& photometry magnitude\\
$w?mag$ & 8.25$^{\prime\prime}$ aperture magnitudes \\ 
$w?sigm$ & 8.25$^{\prime\prime}$ aperture magnitude uncertainties \\ 
$w?flg$ & 8.25$^{\prime\prime}$ measurement quality flag \\
$w?mag\_1$ & 5.5$^{\prime\prime}$ aperture magnitudes \\
$w?sigm\_1$ &  5.5$^{\prime\prime}$ aperture magnitude uncertainties\\
$w?flg\_1$ & 5.5$^{\prime\prime}$ measurement quality flag \\
$w?mag\_4$ & 13.75$^{\prime\prime}$ aperture magnitudes \\
$w?sigm\_4$ & 13.75$^{\prime\prime}$ aperture magnitude uncertainties\\
$w?flg\_4$ & 13.75$^{\prime\prime}$ measurement quality flag \\
$w?mag\_7$ & 22$^{\prime\prime}$ aperture magnitudes \\
$w?sigm\_7$ &  22$^{\prime\prime}$ aperture magnitude uncertainties\\
$w?flg\_7$ & 22$^{\prime\prime}$ measurement quality flag \\
$w?sat$ & Saturated pixel fraction \\
$ext\_flg$ & Extended source flag\\
$ph\_qual$ & Photometric quality flag \\
$cc\_flags$ & Contamination and confusion flags \\
$moon\_lev$ & Scattered moonlight contamination flag \\
\toprule
RACS-mid & \\
\toprule
$Total\_flux$ & Total flux density \\
$E\_Total\_flux$ & Total flux density error \\
$Peak\_flux$ &  Peak flux density \\
$E\_Peak\_flux$ & Peak flux density error \\
$E\_Total\_flux\_PyBDSF$ & Python Blob Detector and Source Finder \vspace{-0.1cm}\\ 
& (PyBDSF, \citet{Mohan2015}) total \vspace{-0.1cm} \\
& flux density error \\
$E\_Peak\_flux\_PyBDSF$ & PyBDSF peak flux density error \\
$Maj\_axis$ & Major axis size \\
$E\_Maj\_axis$ & Major axis size uncertainty \\
$Min\_axis$ & Minor axis size \\
$E\_Min\_axis$ & Minor axis size uncertainty \\
$PA$ & Position angle \\
$E\_PA$ & Position angle uncertainty \\
$DC\_Maj\_axis$ & Deconvolved source major axis size \\
$E\_DC\_Maj\_axis$ & Deconvolved source major axis uncertainty \\
$DC\_min\_axis$ & Deconvolved source minor axis size \\
$E\_DC\_min\_axis$ & Deconvolved source minor axis uncertainty\\
$DC\_PA$ & Deconvolved source position angle\\
$E\_DC\_PA$ &  Deconvolved source position angle uncertainty \\
$PSF\_Maj$ & Point spread function major axis size\\ 
$PSF\_Min$ & Point spread function minor axis size \\ 
$PSF\_PA$ & Point spread function position angle\\
$E\_Flux\_scale$ & Fraction brightness scale uncertainty\\
$S\_Code$ & PyBDSF source classification code \\
$Flag$ & Source type flag \\
$N\_Gaussians$ & Number of Gaussian components \\
$Noise$ & RMS noise \\
\toprule
\end{tabular*}
\label{tab:feature list}
\end{table}

\subsection{Defining Qualitative Flags}\label{sec:feature_engineering}

The WISE catalogue contains three categorical features, $ph\_qual$, $moon\_lev$, and $cc\_flags$, which are all four-character strings, with each character reflecting the status in one of the four WISE passbands. These features were encoded following the method outlined in \cite{Beck2022}. Each of the four-character strings was first split into four individual features. The resulting $ph\_qual$ features were encoded by applying the following map: A$\rightarrow$7, B$\rightarrow$6, C$\rightarrow$4, U$\rightarrow$3, X$\rightarrow$1, and Z$\rightarrow$0, while the following transformation was used for $cc\_flag$ features: 0$\rightarrow$10, o$\rightarrow$8, h$\rightarrow$7, p$\rightarrow$6, d$\rightarrow$5, O$\rightarrow$3, H$\rightarrow$2, P$\rightarrow$1, and D$\rightarrow$0. For $moon\_lev$, no encoding was required as the original four-character string consisted of integers. 

The RACS-mid catalogue contains one categorical feature, $S\_Code$. The two $S\_Code$ values, `S' and `M', indicate whether a given source was decomposed by PyBDSF into single (`S') or multiple (`M') Gaussians. These were encoded using dummy variables, creating two new inputs for the model ($S\_Code\_M$ and $S\_Code\_S$).

Missing values were set to $-999$, and all features were normalised to a value between 0 and 1 using the \texttt{MinMaxScaler} from \texttt{Scikit-learn}. Finally, the original string class labels from the \citet{Beck2022} catalogue (GALAXY or QUASAR) were assigned integer labels of 0 for quasars and 1 for galaxies.

Following feature engineering, 28 RACS-mid features and 77 WISE features were supplied as inputs to the model, giving 105 features in total. 

\subsection{Hyperparameter Optimisation}\label{ref:hyperparams}

Model hyperparameters were optimised and the performance and generalisability of the model was determined using \texttt{cross\_val\_score} from \texttt{Scikit-learn} to perform a stratified five-fold cross validation. Given the large number of hyperparameters available to \texttt{XGBClassifier} models, an intial hyperparameter search was conducted using \texttt{Scikit-learn}'s \texttt{RandomizedSearchCV} over 200 iterations to determine appropriate hyperparameter search ranges for a second stage of optimisation using the Bayesian hyperparameter optimisation package \texttt{Hyperopt} \citep{Bergstra2013}. 

Hyperparameters that were tuned included the maximum tree depth (\texttt{max\_depth} $\epsilon \, [3, 10]$), the minimum child weight (\texttt{min\_child\_weight} $\epsilon \, [1, 10]$), the minimum loss reduction required to add a partition to the tree (\texttt{gamma} $\epsilon \, [0.1, 5]$), the learning rate (\texttt{eta} $\epsilon \, [0.01, 3]$), the subsample ratio for training (\texttt{subsample} $\epsilon \, [0.5,1 ]$), the number of estimators, (\texttt{n\_estimators} $ \epsilon \, [100, 2000]$), the fraction of columns to sample in each tree (\texttt{colsample\_bytree} $ \epsilon \, [0.6, 1]$), the L1 and L2 regularisation (\texttt{lambda, alpha} $\epsilon \, [0, 100]$), and the balance of positive and negative weights (\texttt{scale\_pos\_weight} $\epsilon \, [0, 10]$).

To prevent data leakage, a pipeline was used to resample and train the model for each cross-validation evaluation (including within hyperparameter search commands) to ensure the holdout fold was not inadvertently resampled. The final model hyperparameters are listed in Table \ref{tab:optimised_params}.

\begin{table}[]
\centering
\caption{Final optimised hyperparameters for the \texttt{XGBClassifier} model with random oversampling of the quasar class.}
\begin{tabular*}{0.48\textwidth}{@{\extracolsep{\fill}} l c }
\toprule \toprule
Parameter & Value \\
\toprule
max\_depth & 7 \\
n\_estimators & 706 \\
learning\_rate & 0.03 \\
colsample\_bytree & 0.61 \\ 
min\_child\_weight & 2 \\
scale\_pos\_weight & 5 \\
subsample & 0.53 \\
gamma & 1.80 \\
alpha & 0 \\
lambda & 0.1 \\
early\_stopping\_rounds & 100 \\
\toprule
\end{tabular*}
\label{tab:optimised_params}
\end{table}

\section{Understanding the Model}

\renewcommand{\thefigure}{B\arabic{figure}}
\setcounter{figure}{0}

\renewcommand{\thetable}{B\arabic{table}}
\setcounter{table}{0}

\subsection{Redshift Distribution}\label{sec:redshift_effects}

The redshift distribution of the sources classified as galaxies by the optimised model is shown as the blue line in Figure \ref{fig:z_dist}. Galaxy redshifts were obtained from the dense neural network predictions of \citet{Beck2022}, however, no redshift information is available for sources they predicted to be quasars.  The distribution is characterised by a mean of 0.53 and a standard deviation of 0.30. $\sim96\%$ have redshifts less than 1, and $\sim 9\%$ have redshifts less than 0.1. Given the shallow nature of RACS-mid, a large fraction of high-redshift sources will be missed. As such, high completeness is only expected for $z<0.1$. Considering only galaxies with $z<0.1$, the mean redshift and standard deviation are 0.06 and 0.02, respectively.

For comparison, the dotted red line in Figure \ref{fig:z_dist} shows the redshift distribution for galaxies that were incorrectly classified as quasars by the model. Since this distribution peaks at $z\sim0.9$, it appears that the model struggles to classify high-redshift sources, and often confuses galaxies and quasars at $z\gtrsim0.5$. This may be related to the fact that observations of high-redshift sources are noisier, reducing model accuracy, an effect which is likely amplified by the short integration time of RACS. Since the goal of RACS-mid was to rapidly survey the full radio sky, deeper observations at 1.4 GHz will enable better completeness across all redshifts, allowing future machine-learning models similar to the one implemented in this paper to more accurately predict the nature of high-redshift sources.

\begin{figure}
    \centering
    \includegraphics[width=0.99\linewidth]{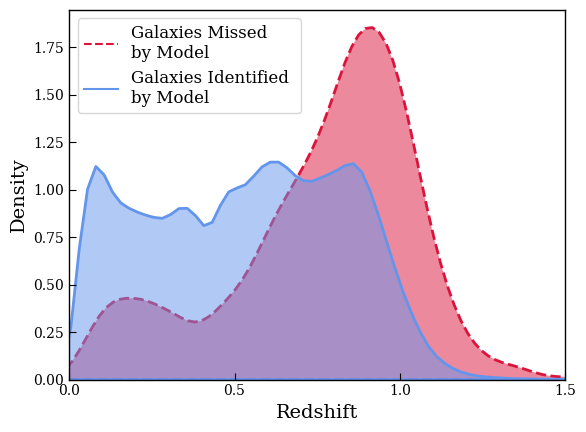}
    \caption{The redshift distribution of the sources classified as galaxies by the final model (blue line), compared to that of the galaxies that were missed by the model i.e. were incorrectly classified as quasars (dotted red line). Note that these distributions are kernel density estimates, and therefore the probability is given by the area under the curve, which sums to one.}
    \label{fig:z_dist}
\end{figure}

\subsection{The Effects of Including RACS-mid Features on Model Performance} \label{sec:shap}

To determine the effects of including radio continuum features in the classifier model, the same model parameters and oversampling technique as outlined in Section \ref{sec:results} were used, but only WISE or RACS-mid features were selected as inputs for the training data set, resulting in 77 or 28 input parameters, respectively. After training the model with these new features, model performance was assessed using the same evaluation metrics discussed in Section \ref{sec:results}, which are presented in Table \ref{tab:feature_trial_metrics}.

\begin{table}[]
\label{tab:feature_trial_metrics}
\centering
\caption{Performance metrics for \texttt{XGBClassifier} models trained on either WISE or RACS-mid features only.}
\begin{tabular*}{0.48\textwidth}{@{\extracolsep{\fill}} l c c}
\toprule \toprule
Metric & Value  & Value \vspace{-0.1cm}\\
 & (WISE only) & (RACS-mid only) \\
\toprule
accuracy & 0.93 & 0.86 \\
weighted F1-score & 0.93 & 0.81 \\
quasar precision & 0.79 & 0.58 \\
quasar recall & 0.73 & 0.09 \\
quasar F1-score & 0.76 &  0.16 \\
galaxy precision & 0.96& 0.87 \\
galaxy recall &  0.97 & 0.99 \\
galaxy F1-score & 0.96 & 0.92 \\
$\kappa$ & 0.72 & 0.12 \\
cross-entropy & 0.16 & 0.37 \\
ROC AUC & 0.96 & 0.70 \\
AUPRC & 0.99 & 0.93\\
\toprule
\end{tabular*}

\end{table}

From the WISE-only results, it is evident that the addition of RACS-mid features provides a minor improvement to model performance due to a reduction in the rate of false positives and negatives in the test data set. This results in  minor increases in accuracy, quasar precision, $\kappa$, and the area under the ROC curve compared to a model trained on WISE features alone. 

\begin{figure*}
    \centering
    \includegraphics[width=0.47\linewidth]{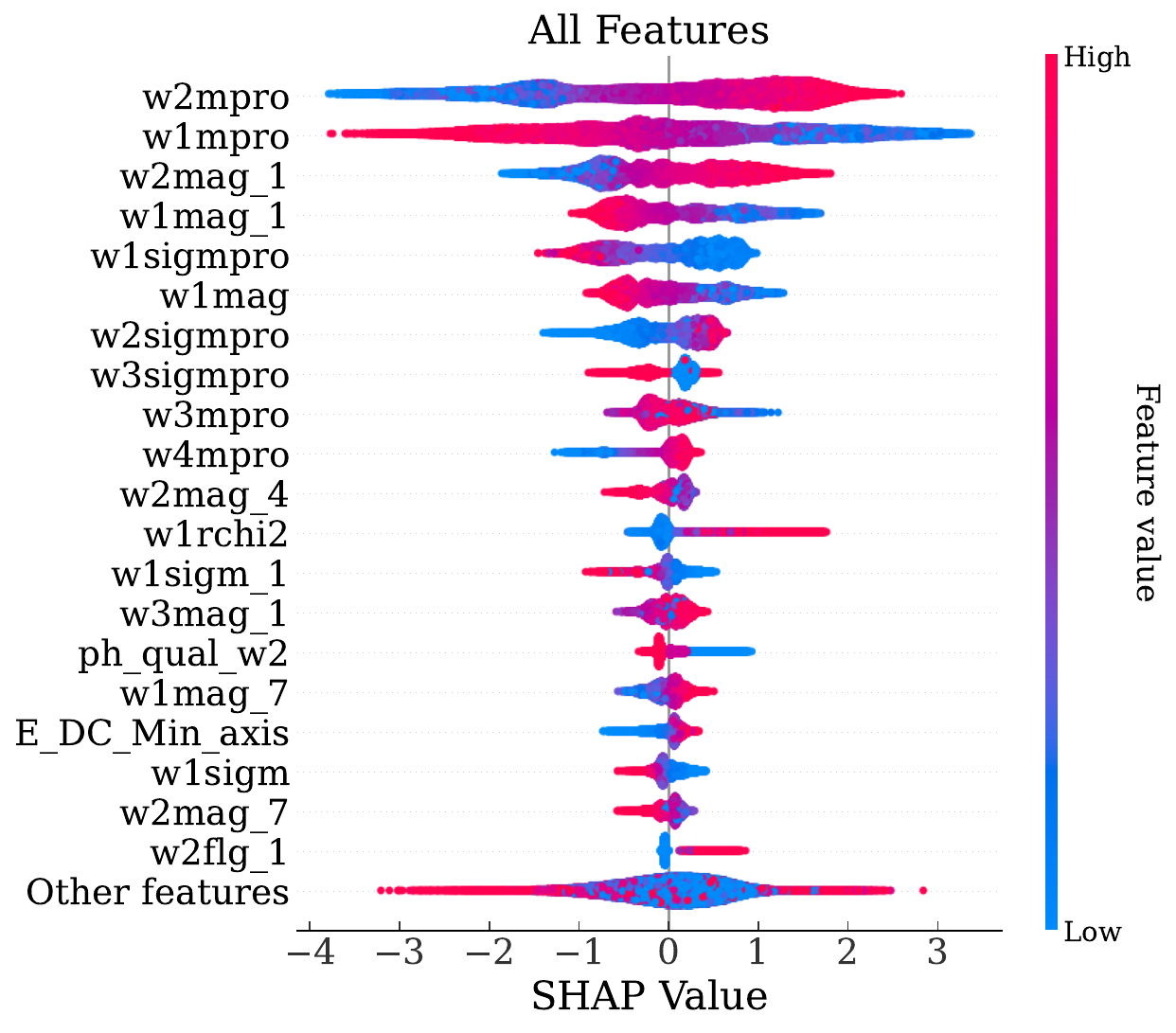}
    \includegraphics[width=0.5\linewidth]{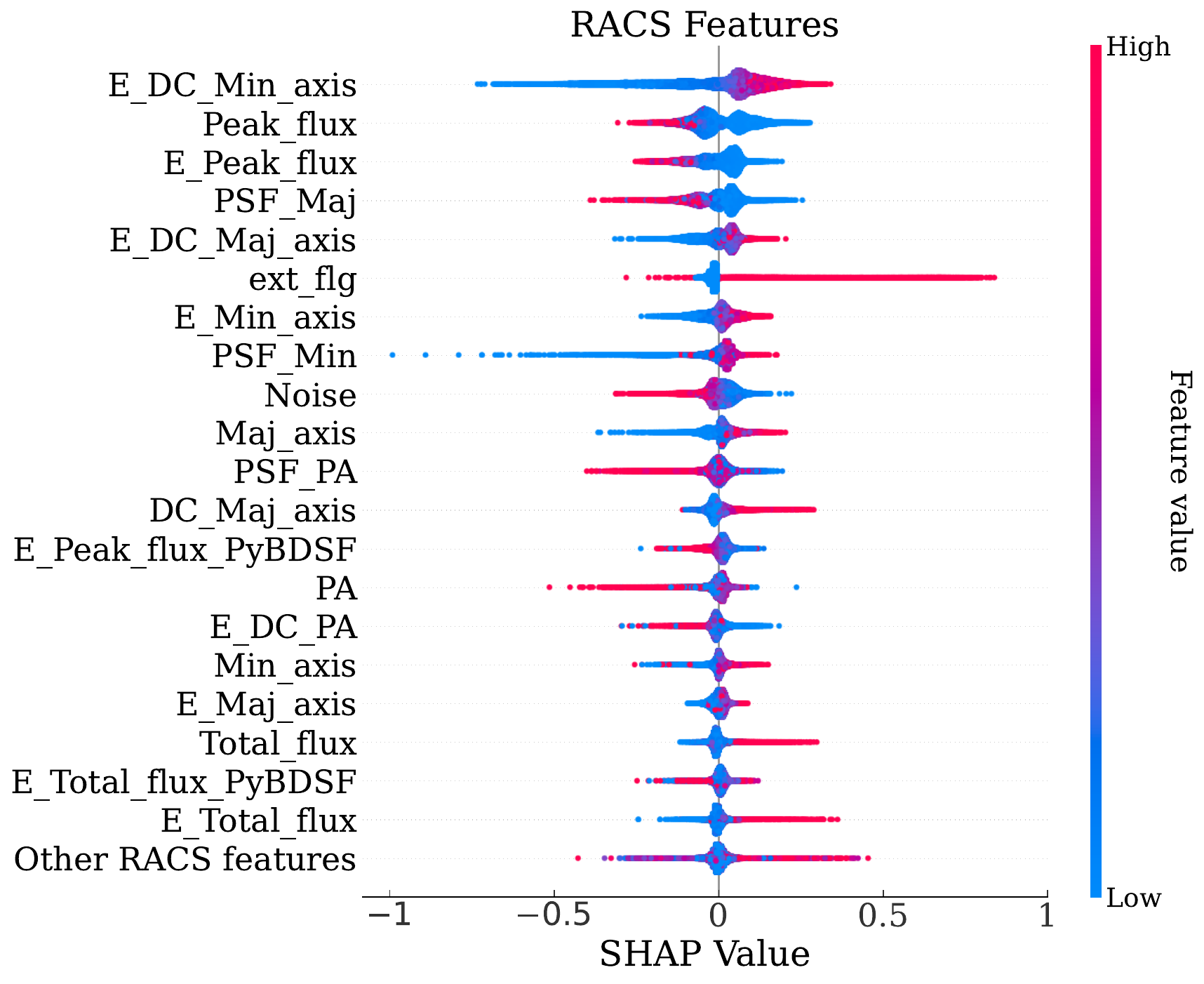}
    \caption{SHAP analysis results for the top 20 model features, including both WISE and RACS-mid features (left panel) and RACS-mid features only (right panel). For each of the displayed features, a data point represents whether that feature had a positive (SHAP values $>1$) or negative (SHAP values $<1$) influence on the prediction as a galaxy. The colourbar indicates whether a feature value was high or low compared to the population.}
    \label{fig:shap_analysis_all}
\end{figure*}

This is consistent with the results of the Shapley Additive Explanations (SHAP) analysis \citep{Lundberg2017} that was performed on the optimised \texttt{XGBClassifier} model using all 105 RACS-mid and WISE features. The distribution of SHAP values for the top 20 model features and RACS-mid features are shown in the left and right panels of Figure \ref{fig:shap_analysis_all}, respectively. Positive SHAP values indicate a feature had a positive impact on the prediction (galaxy classification), and vice versa. For each feature, a data point is added indicating the impact on the model prediction, with the colour bar showing whether the feature value was low (blue) or high (red) compared to the population. For example, from the left panel of Figure \ref{fig:shap_analysis_all}, high values of $w2mpro$ (the W2-band magnitude measured using profile-fitting photometry) made a galaxy classification more likely, while low values were associated with quasar predictions. However, for $w1mpro$, this trend is reversed. SHAP values also indicate the magnitude of the impact a given feature has on the model's predictions. For instance, $w4mpro$ exhibits a similar overall behaviour to $w2mpro$, with higher values having a positive impact on model predictions and low values having a negative impact. However, the magnitudes of the SHAP values are larger for $w2mpro$, and thus it tends to have a greater influence on the model's output compared to $w4mpro$.

From the left panel of Figure \ref{fig:shap_analysis_all}, it is clear that WISE features have the greatest effect on the model predictions, making up 18 of the top 19 most influential features. In fact, the only RACS-mid feature in the top 19 features is \textit{E\_DC\_Min\_axis}, the uncertainty in the deconvolved source minor axis, with high values having a positive impact on model performance, albeit to a lesser extent than some of the WISE features. This is consistent with the small ($\sim1\%$) decrease in model performance that is associated with the removal of RACS-mid features from the training data set. Since the SHAP analysis indicates that WISE features have the largest influence on model predictions, it is not surprising that the effects of removing RACS-mid data from the model inputs are minor.  

The top features in the left panel of Figure \ref{fig:shap_analysis_all} indicate that quantities associated with the WISE W1 and W2 bands are driving factors for the model's predictions. This is in line with other methods used to separate populations of galaxies and quasars, which commonly rely on IR colours, with star-forming galaxies characterised by bluer W1 - W2 colours than quasars \citep{Lacy2004, Sajina2005, Stern2005, Lacy2007, Messias2012, Messias2014}. 

Considering the right panel of Figure \ref{fig:shap_analysis_all}, the most important RACS-mid features for the model were the values and uncertainty associated with the source and deconvolved source major and minor axes (\textit{E\_DC\_Min\_axis, E\_DC\_Maj\_axis, E\_Min\_axis, Maj\_axis, DC\_Maj\_axis, Min\_axis, E\_Maj\_axis}), the values and uncertainty in the flux and peak flux (\textit{Peak\_Flux, E\_Peak\_Flux, E\_Peak\_Flux\_PyBDSF, Total\_Flux, E\_Total\_Flux\_PyBDSF, E\_Total\_Flux}), the major and minor axes and position angle of the point-spread function of the field (\textit{PSF\_Maj, PSF\_Min, PSF\_PA}), and the quantities and uncertainties associated with the position angle of the source (\textit{PA, E\_DC\_PA}), and the rms noise. Thus, it appears that the model tends to classify larger objects with greater total fluxes but lower peak fluxes as galaxies i.e. extended sources that may not have a significantly brighter central source such as an accreting supermassive black hole at the centre.

In addition to the model trained on WISE features alone, another model was trained solely on RACS-mid features to determine whether radio-continuum 1.4 GHz observations contain sufficient information to discriminate between populations of galaxies and quasars. The performance of this model is summarised in Table \ref{tab:feature_trial_metrics}. While the accuracy remains high ($\sim86\%)$, this is not surprising or particularly informative given that the rate of occurrence of galaxies is also $\sim86\%$. In effect, the model appears to be assigning galaxy classifications to most objects, consistent with the low recall score for quasars of 0.09. Therefore, if a high rate of quasar contamination is not a deterring factor, this result suggests that machine-learning models trained on 1.4 GHz radio-continuum data alone can be a cheap and efficient method to obtain a relatively pure (precision of 0.87) population of SFGs. The use of both IR and radio-continuum features significantly enhances model performance, with 8\% and 12\% increases in the model accuracy and weighted F1 score, respectively. The greatest improvements associated with the introduction of WISE features to the model are quasar precision and recall, which increase from 0.58 to 0.80, and 0.09 to 0.73, respectively. However, given there is evidence to suggest that the IRRC is non-linear or redshift-dependent \citep[e.g.,][]{Ivison2010, Sargent2010, Smith2014b, Molnar2021, McCheyne2022}, this behaviour may change at higher redshifts.

\subsection{IR Colours of the Galaxy and Quasar Populations}

Figure \ref{fig:ir_colours_all} shows the WISE IR colours $W1- W2$ and $W2 - W3$ for all sources that were classified by the model and for those sources that were incorrectly classified in the left and right panels, respectively. The colour map indicates the ratio of predicted galaxies to quasars in each pixel, and the mid-IR colour distributions for the populations of galaxies (purple), $z<0.1$ galaxies (dotted pink), and quasars (blue) are shown alongside each axis. Note that the right panel of Figure \ref{fig:ir_colours_all} does not show the distributions for the $z<0.1$ incorrect galaxies (which are actually quasars), as the \citet{Beck2022} catalogue only provides redshifts for sources they identified as galaxies.

The WISE colour-colour plane is often divided into different regions based on the source of the observed IR emission. Sources whose mid-IR emission is dominated by heating from dust-obscured AGN occupy the top of the graph ($W1-W2 > 0.8$). For those sources whose mid-IR colours do not indicate significant AGN activity, the plane may be further subdivided based on the $W2 - W3$ colour, which is a proxy for the SFR. Early-type, low-SFR galaxies occupy the bottom left of the plane ($W2-W3<2$), spiral galaxies are in the centre ($2<W2-W3<3.5$), and starburst galaxies occupy the bottom right ($W2-W3>3.5$) \citep{Jarrett2011, Jarrett2013, Cluver2014, Grundy2023}.

\begin{figure*}
    \centering
    \includegraphics[width=0.49\linewidth]{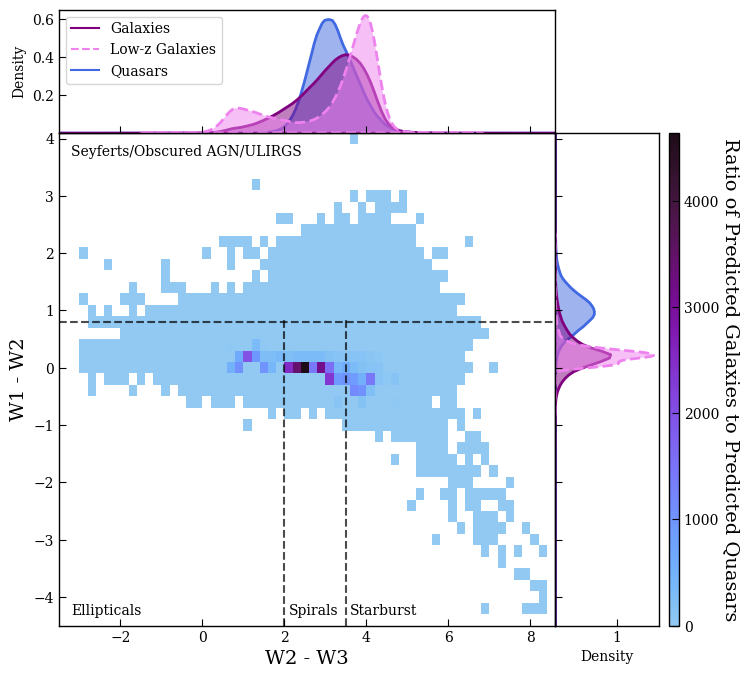}
    \includegraphics[width=0.48\linewidth]{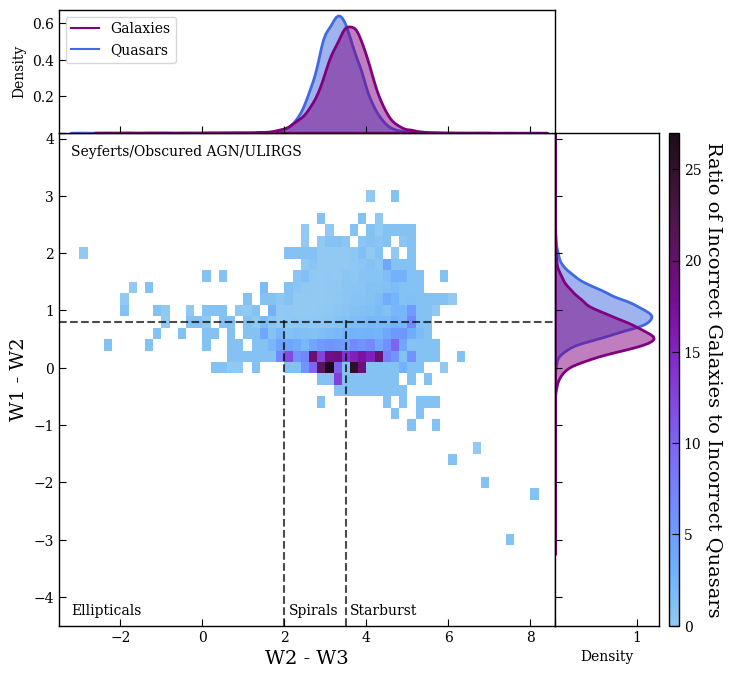} 
    \caption{WISE IR colours of all model predictions (left panel) and for the incorrect model prediction (right panel). The colour map indicates the ratio of galaxies to quasars in each region of the figure, and the mid-IR colour distributions are shown for galaxies (purple), quasars (blue), and $z<0.1$ galaxies (dotted pink) beside each axis.}
    \label{fig:ir_colours_all}
\end{figure*}

From the left panel of Figure \ref{fig:ir_colours_all}, the ratio of galaxies to quasars is highest in the centre of the graph, where spiral galaxies, as well as elliptical and starburst galaxies, are expected to lie. There is also a tail that extends to the bottom right of the graph, which indicates the presence of a population of high-SFR starburst galaxies. Considering the distribution on the y-axis, there is a clear distinction between the $W1-W2$ colour distributions for quasars compared to galaxies, consistent with the expectation that AGN are characterised by redder $W1-W2$ colours than galaxies \citep{Jarrett2011, Jarrett2013, Cluver2014, Grundy2023} and the SHAP results in Section \ref{sec:shap} that indicate that high $W2$ and low $W1$ values (i.e. low $W1-W2$) had a strong influence in favour of a galaxy prediction. This is in contrast to the overlapping $W1-W2$ distributions for the incorrectly classified galaxies and quasars in the right panel of Figure \ref{fig:ir_colours_all}. Based on this, the model appears to struggle with sources whose $W1-W2$ colours are located near the boundary at $W1-W2=0.8$. Such sources may exhibit properties associated with both galaxies and quasars, for example the presence of an actively accreting AGN as well as significant star-formation activity.

The $W2-W3$ distributions in the left panel of Figure \ref{fig:ir_colours_all} overlap significantly, although the tail of the galaxy population, largely comprising very low-redshift sources ($z\lesssim0.1)$, does extend to lower $W2-W3$ values. This is consistent with previous studies that did not find WISE $W2-W3$ colours to be a useful indicator of AGN activity \citep{Jarrett2011, Jarrett2013, Cluver2014, Grundy2023}. Moreover, the SHAP results in Section \ref{sec:shap} indicate that features associated with the $W3$ band have a smaller influence on model predictions than W1- and W2-band features. However, low $w3mpro$ values have a strong influence on galaxy predictions, an effect that may be related to the tail in the $W2-W3$ galaxy distribution. Compared to the $W2-W3$ distributions for all predictions in the top panel, the $W2-W3$ distributions of the incorrect galaxies and quasars in the bottom panel are even more similar. The tail of galaxies with low $W2-W3$ values has disappeared, indicating that the model was in general able to correctly classify those sources. Thus, the model appears to make incorrect predictions for sources whose $W1-W2$ colours lie close to the boundary at $0.8$, and/or whose $W2-W3$ colours are consistent with either population.
\end{appendix}

\end{document}